\documentclass[useAMS,usenatbib]{mn2e}
\newcommand{\parcsec}{\mbox{$\stackrel{\prime\prime}{\textstyle .}$}}
\newcommand{\parcmin}{\mbox{$\stackrel{\prime}{\textstyle .}$}}
\newcommand{\nitrogen}{[\mbox{N\,{\sc ii}}]}
\makeatletter

\newcommand{\Rmnum}[1]{\expandafter\@slowromancap\romannumeral #1@}
\makeatother
\usepackage{graphicx,amsmath,amssymb,bm}
 \usepackage{color,epstopdf}
\newcommand{\dtm}{$\mathcal{{DT\!\!M}}$}
\usepackage{array}
\usepackage{url}

\title[GRB\,121024A]{The warm, the excited, and the molecular gas: \\ 
GRB\,121024A shining through its star-forming galaxy.  \thanks{Based on observations
carried out under prog. ID 090.A-0088(B) with the X-shooter spectrograph
installed at the Cassegrain focus of the Very Large Telescope (VLT), Unit 2 --
Kueyen, operated by the European Southern Observatory (ESO) on Cerro Paranal,
Chile. Also used are observations made with the Nordic Optical Telescope, operated
by the Nordic Optical Telescope Scientific Association, and the Gran Telescopio Canarias (program GTC67-13B), both at the Observatorio del
Roque de los Muchachos, La Palma, Spain, of the Instituto de Astrofisica de
Canarias. HAWK-I imaging used is part of the program 092.A-0076(B).
This work made use of data supplied by the UK Swift Science Data Centre at the University of Leicester.}}
\author[M. Friis et al.]
  {M.~Friis$^1$,
 A.~De Cia$^2$, T.~Kr\"{u}hler$^{3,4}$, J. P. U.~Fynbo$^4$, C.~Ledoux$^3$, P. M.~Vreeswijk$^2$, 
 \newauthor 
D. J.~Watson$^4$, D.~Malesani$^4$, J.~Gorosabel$^{5,6,7}$\textdagger, R. L. C.~Starling$^8$, P.~Jakobsson$^1$,
\newauthor 
K.~Varela$^9$, K.~Wiersema$^8$, A. P.~Drachmann$^{4}$, A. Trotter$^{10,11}$, C. C.~Th\"{o}ne$^5$,
\newauthor
A.~de Ugarte Postigo$^{5,4}$, V.~D'Elia$^{12,13}$, J.~Elliott$^9$, M.~Maturi$^{14}$, P.~Goldoni$^{15}$,
\newauthor
J.~Greiner$^9$, J.~Haislip$^{10}$, L.~Kaper$^{16}$, F.~Knust$^9$, A.~LaCluyze$^{10}$, B.~Milvang-Jensen$^4$,
\newauthor 
D.~Reichart$^{10}$, S.~Schulze$^{17,18}$, V. Sudilovsky$^9$, N. Tanvir$^{8}$, S. D.~Vergani$^{19}$ \\
  \textdagger Deceased \\
  $^1$Centre for Astrophysics and Cosmology, Science Institute, University of Iceland, Dunhagi 5, 107 Reykjav\'ik, Iceland \\
  $^2$Department of Particle Physics and Astrophysics, Faculty of Physics, Weizmann Institute of Science, 76100, Rehovot, Israel \\
  $^3$European Southern Observatory, Alonso de C\'ordova 3107, Casilla 19001, Santiago 19, Chile \\
  $^4$Dark Cosmology Centre, Niels Bohr Institute, University of Copenhagen, Juliane Maries Vej 30, 2100 Copenhagen, Denmark \\
  $^5$Instituto de Astrof\'isica de Andaluc\'ia (IAA-CSIC), Glorieta de la Astronom\'ia s/n, 18008 Granada, Spain \\
  $^6$Unidad Asociada Grupo Ciencia Planetarias UPV/EHU-IAA/CSIC, Departamento de F\'isica Aplicada I, E.T.S. Ingenieria,  \\ Universidad del Pais Vasco UPV/EHU, Alameda de Urquijo s/n, 48013 Bilbao, Spain \\
  $^7$Ikerbasque, Basque Foundation for Science, Alameda de Urquijo 36-5, 48008 Bilbao, Spain \\
  $^8$Department of Physics and Astronomy, University of Leicester, University Road, Leicester LE1 7RH, UK \\
  $^9$Max-Planck-Institut f\"{u}r extraterrestrische Physik, Giessenbachstra\ss e 1, 85748 Garching, Germany \\
  $^{10}$Department of Physics and Astronomy, University of North Carolina at Chapel Hill, Campus Box 3255, Chapel Hill, NC 27599, USA \\
  $^{11}$Department of Physics, NC A\&T State University, 1601 E. Market St, Greensboro, NC 27411, USA \\
  $^{12}$INAF/Rome Astronomical Observatory, via Frascati 33, I-00040 Monteporzio Catone (Roma), Italy \\
  $^{13}$ASI-Science Data Center, Via del Politecnico snc, I-00133 Rome, Italy \\
  $^{14}$Zentrum f\"{u}r Astronomie der Universit\"{a}t Heidelberg, Institut f\"{u}r Theoretische Astrophysik, Philosophenweg 12, 69120 Heidelberg, Germany \\
  $^{15}$APC, Astroparticule et Cosmologie, Universite Paris Diderot, CNRS/IN2P3, CEA/Irfu, Observatoire de Paris, Sorbonne Paris Cit\'{e}, \\ 10 Rue Alice Domon et Leonie Duquet, F-75205 Paris, Cedex 13, France \\
  $^{16}$Anton Pannekoek Institute for Astronomy, University of Amsterdam, Science Park 904, 1098 XH, Amsterdam, The Netherlands \\
  $^{17}$Millennium Institute of Astrophysics, Casilla 306, Santiago 22, Chile \\
  $^{18}$Instituto de Astrof\'isica, Facultad de F\'isica, Pontif\'icia Universidad Catolica de Chile, Casilla 306, Santiago 22, Chile \\
  $^{19}$Laboratoire GEPI, Observatoire de Paris, CNRS-UMR8111, Universit\'{e} Paris Diderot, 5 place Jules Janssen, F-92195 Meudon, France}
\date{}

\pagerange{\pageref{firstpage}--\pageref{lastpage}} \pubyear{}

\def\LaTeX{L\kern-.36em\raise.3ex\hbox{a}\kern-.15em
    T\kern-.1667em\lower.7ex\hbox{E}\kern-.125emX}

\begin{document}

\label{firstpage}

\maketitle
 
\clearpage
 
\begin{abstract}
We present the first reported case of the simultaneous metallicity determination
of a gamma-ray burst (GRB) host galaxy, from both afterglow absorption lines as well as
strong emission-line diagnostics. Using spectroscopic and imaging observations of the afterglow and
host of the long-duration \emph{Swift} GRB\,121024A at $z\,=\,2.30$, we give
one of the most complete views of a GRB host/environment to date. We observe a
strong damped Ly$\alpha$ absorber (DLA) with a hydrogen column density of log\,$N(\text{H\,{\sc
i}})\,=\,21.88\pm0.10$, H$_2$ absorption in the Lyman-Werner bands
(molecular fraction of log($f$)\,$\approx-1.4$; fourth solid detection of molecular hydrogen in a GRB-DLA), 
the nebular emission lines H$\alpha$, H$\beta$, [\mbox{O\,{\sc ii}}], [\mbox{O\,{\sc iii}}] and [\mbox{N\,{\sc ii}}], as well as metal absorption lines. 
We find a GRB host galaxy that is highly star-forming (SFR\,$\sim$\,40\,M$_\odot$\,yr$^{-1}$), with a dust-corrected metallicity along the line of sight of
[Zn/H]$_{\rm corr} =-0.6\pm0.2$ ($\text{[O/H]}\sim-0.3$ from emission lines), and a depletion factor \lbrack Zn/Fe\rbrack\,=\,$0.85\pm0.04$. 
The molecular gas is separated by 400\,km\,s$^{-1}$ (and 1--3\,kpc) from the gas that is photo-excited by the GRB. This implies a fairly massive host, in agreement with the derived stellar mass of log(M$_*$/M$_\odot$) = $9.9^{+0.2}_{-0.3}$. 
We dissect the host galaxy by characterising its molecular component, the excited gas, and the line-emitting star-forming regions. 
The extinction curve for the line of sight is found to be unusually flat ($R_V\sim15$). We discuss the possibility of an anomalous grain size distributions.
We furthermore discuss the different metallicity determinations from both absorption and emission lines, which gives consistent results for the line of sight to GRB\,121024A.

\end{abstract}


\section{Introduction}

The study of gamma-ray burst (GRB) afterglows has proven to be a powerful tool
for detailed studies of the interstellar medium (ISM) of star-forming galaxies,
out to high redshifts
\citep[e.g.][]{vreeswijk04,prochaska07,ledoux09,sparre13}. With quickly fading
emission spanning the entire electromagnetic spectrum, GRB afterglows offer
a unique opportunity to probe the surrounding environment. The intrinsic
spectrum of the afterglow is well fitted with simple power-law segments, so the
imprints of the intergalactic medium (IGM) as well as the ISM surrounding the burst
are relatively easy to distinguish from the afterglow in the observed spectrum. Moreover, with absorption and emission-line analysis it 
is possible to determineparameters such as H\,{\sc i} column density, metallicity, dust
depletion, star-formation rate (SFR) and kinematics of the GRB host galaxy.


Metallicity is a fundamental parameter for characterising a galaxy and it
holds important information about its history. Metallicity might also play a crucial role in 
the GRB production mechanism. For GRB hosts, the metallicity is
measured either from hydrogen and metal absorption lines, or by using
diagnostics based on the fluxes of strong nebular emission lines, calibrated in
the local Universe. Different calibrations are in use leading to some
discrepancy \citep[e.g.][]{Kudritzki}, and the different diagnostics have
their strengths and weaknesses (e.g. less sensitive to reddening, multiple solutions, 
or more sensitive at high metallicities). 
The absorption lines probe the ISM along the line of sight,
while the nebular line diagnostics determine the integrated metallicity of the H\,{\sc ii} regions of the
host. For GRB damped Ly$\alpha$ absorbers \citep[GRB-DLAs, $N$(H\,{\sc i})$>2\times10^{20}$\,cm$^{-2}$][]{wolfe05}, a direct comparison of
metallicity from the two methods is interesting because it can either provide a 
test of the strong-line methods or alternatively allow a measurement of a possible
offset in abundances in H\,{\sc ii} regions and in the ISM. So far, this 
comparison has only been carried out for a few galaxy counterparts of DLAs found in the line of sight of background QSOs 
\citep[QSO-DLAs, e.g.][]{bowen05,peroux,noterdaeme12,fynbo13,JW14}. To our knowledge, a comparison for GRB-DLAs has
 not been reported before. For both emission and absorption measurements to be feasible with current instrumentation, the observed
host needs to be highly star-forming, to have strong nebular lines, and at the
same time be at a redshift high enough for the Ly$\alpha$ transition to be
observed (at redshifts higher than $z\approx1.5$ the
Ly$\alpha$ absorption line is redshifted into the atmospheric transmission
window). GRB\,121024A is a $z=2.30$ burst hosted by a highly star-forming galaxy.
We measure abundances of the GRB host galaxy in absorption and compare them with the metallicity determined by
strong-line diagnostics using observed nebular lines from [\mbox{O\,{\sc ii}}],
[\mbox{O\,{\sc iii}}], [\mbox{N\,{\sc ii}}] and the Balmer emission lines.


Apart from the absorption features from metal lines, we also detect the
Lyman-Werner bands of molecular hydrogen. Molecular hydrogen is hard to detect in absorption, because it requires high S/N and mid-high resolution. 
As long duration GRBs (t$_{\text{obs}}>\,$2\,s) are thought to be associated with the death of
massive stars \citep[e.g.][]{hjorth03,stanek03,sparre11,cano13,schulze14}, they are
expected to be found near regions of active star formation, and hence molecular
clouds. In spite of this, there are very few detections of molecular absorption
towards GRBs  \citep[see e.\,g][]{tumlinson07}. \cite{ledoux09} found that this is likely due to the low
metallicities found in the systems observed with high resolution spectrographs
($R=\lambda/\Delta\lambda\gtrsim40000$). Typically, mid/high-resolution spectroscopy at a sufficient S/N is only possible for the brighter sources.
As is the case for QSO-DLAs, lines of sight with
H$_2$ detections will preferentially be metal-rich and dusty. The observed spectra are therefore
UV-faint and difficult to observe \citep[GRB\,080607 is a striking exception, where observations were possible 
thanks to its extraordinarily intrinsic luminosity and 
rapid spectroscopy, see][]{prochaska09}. Now with X-shooter \citep{xshooter}
on the Very Large Telescope (VLT) we are starting to secure spectra with 
sufficient resolution to detect H$_2$ for fainter systems resulting in additional detections \citep{thomas1,delia14}.

Throughout this paper we adopt a flat $\Lambda$CDM cosmology with $H_0\,=\,71$\,km\,s$^{-1}$ and $\Omega_\text{M}\,=\,0.27$, and report $1\,\sigma$ errors ($3\,\sigma$ limits), unless otherwise indicated. Reference solar abundances are taken from \cite{asplund09}, where either photospheric or meteoritic values (or their average) are chosen according to the recommendations of \cite{lodders09}. Column densities are in cm$^{-2}$. In Sect.~\ref{obs} we describe the data and data reduction used in this paper, in Sect.~\ref{results} we present the data analysis and results, which are then discussed in Sect.~\ref{discussion}.

\section{Observations and Data Reduction}\label{obs}
On 2012 October 24 at 02:56:12 UT the Burst Alert Telescope \citep[BAT,][]{bat} onboard the \emph{Swift} satellite \citep{swift} triggered on GRB\,121024A. The X-Ray Telescope (XRT) started observing the field at 02:57:45 UT, 93 seconds after the BAT trigger. About one minute after the trigger, Skynet observed the field with the PROMPT telescopes located at CTIO in Chile and the $16"$ Dolomites Astronomical Observatory telescope (DAO) in Italy \citep{prompt} in filters $g'$,\,$r'$,\,$i'$,\,$z'$ and $BRi$. Approximately 1.8 hours later, spectroscopic afterglow measurements in the wavelength range of 3000\,\AA \ to 25\,000\,\AA \ were acquired (at 04:45 UT), using the cross-dispersed, echelle spectrograph X-shooter \citep{xshooter} mounted at ESO's VLT. Then at 05:53 UT, 3 hours after the burst, the Gamma-Ray burst Optical/NIR Detector \citep[GROND,][]{grond1,grond2} mounted on the 2.2 m MPG/ESO telescope at La Silla Observatory (Chile), performed follow-up optical/NIR photometry simultaneously in $g', r', i', z'$ and $JHK$. About one year later (2013 November 07), VLT/HAWK-I imaging of the host was acquired in the $J$ (07:02:13 UT) and $K$ (06:06:47 UT) band. To supplement these, $B$, $R$ and $i$ band imaging was obtained at the Nordic Optical Telescope (NOT) at 2014 January 06 ($i$) and February 10 ($R$) and 19 ($B$). Gran Telescopio Canarias (GTC) observations in the $g$ and $z$ band were optioned on 2014 February 28. For an overview see Tables~\ref{tab:xshooter},~\ref{tab:res} and~\ref{tab:phot}. Linear and circular polarisation measurements for the optical afterglow of GRB\,121024A have been reported in \cite{wiersema}.

\begin{table}
\caption{X-shooter observations}
\renewcommand*{\arraystretch}{1.3}
\begin{tabular}{@{} l c c c c c @{}}
\hline\hline
$t_{\rm obs}$ (UT)$^a$	&	$t_{\rm GRB}$ (min)$^b$	&	$t_{\rm exp.}$ (s)&	Mean Airmass	&	Seeing				\\ \hline
04:47:01				&	116					&	600			&	1.23			&	0\parcsec6-0\parcsec7	\\
04:58:35				&	127					&	600			&	1.19			&	0\parcsec6-0\parcsec7	\\
05:10:12				&	139					&	600			&	1.16			&	0\parcsec6-0\parcsec7	\\
05:21:46				&	151					&	600			&	1.13			&	0\parcsec6-0\parcsec7	\\ \hline
\end{tabular}
\\
\flushleft{
$^a$ Start time of observation on October 24, 2012. \\
$^b$ Mid-exposure time in seconds since GRB trigger.}
\label{tab:xshooter}
\end{table}

\subsection{X-shooter NIR/Optical/UV Spectroscopy}
The X-shooter observation consists of four nodded exposures with exposure times of 600\,s each, taken simultaneously by the ultraviolet/blue (UVB), visible (VIS) and near-infrared (NIR) arms. The average airmass was 1.18 with a median seeing of $\sim$0\parcsec7. The spectroscopy was performed with slit-widths of 1\parcsec0, 0\parcsec9 and 0\parcsec9 in the UVB, VIS and NIR arms, respectively. The resolving power $R=\lambda$/$\Delta$$\lambda$ is determined from telluric lines to be $R\,=\,13000$ for the VIS arm. This is better than the nominal value due to the very good seeing. Following \cite{fynbo11} we then infer $R\,=\,7100$ and $R\,=\,6800$ for the UVB and NIR arms, see Table~\ref{tab:res} for an overview.

X-shooter data were reduced with the ESO/X-shooter pipeline version 2.2.0 \citep{pipeline}, rectifying the data on an output grid with a dispersion of 0.15\,\AA/pixel in the UVB, 0.13\,\AA/pixel in the VIS and 0.5\,\AA/pixel in the NIR arm. The wavelength solution was obtained against arc-lamp frames in each arm. Flux-calibration was performed against the spectrophotometric standard GD71 observed during the same night. 
We further correct the flux-calibrated spectra for slit-losses by integrating over filter curves from GROND photometry shifted to X-shooter observation times (assuming a slope of $\alpha=0.8$). For the UVB arm, only the $g'$ band photometry is available, which covers the DLA (see Sect.~\ref{abs}), making this calibration less secure. Wavelengths are plotted in vacuum and corrected for heliocentric motion.

\begin{table}
\caption{X-shooter resolution}
\renewcommand*{\arraystretch}{1.3}
\begin{tabular}{@{} l c c @{}}
\hline\hline
Arm		&	Slit			&	R\,=\,$\lambda$/$\Delta$$\lambda$	\\ \hline
NIR		&	0\parcsec9	&	6800							\\
VIS		&	0\parcsec9	&	13000						\\
UVB		&	1\parcsec0	&	7100							\\
\end{tabular}
\label{tab:res}
\end{table}

\begin{table*}
\caption{Photometric Observations}
\renewcommand*{\arraystretch}{1.3}
\begin{tabular}{@{} p{3.2cm} >{\centering\arraybackslash}p{2.0cm} >{\centering\arraybackslash}p{2.2cm} >{\centering\arraybackslash}p{2.2cm} >{\centering\arraybackslash}p{2.2cm} >{\centering\arraybackslash}p{2.2cm} @{}}
\hline\hline
Instrument 				&	Time$^{a}$	&	Filter					&	Exp. time (s)		&		Seeing			&	Mag. (Vega)		\\ \hline
MPG/GROND 				&	3.0 h		 	&	$g'$					&	284				&	$1\parcsec55$			&	$20.79\pm0.07$	\\ 
MPG/GROND 				&	3.0 h		 	&	$r'$					&	284				&	$1\parcsec40$			&	$19.53\pm0.05$	\\ 
MPG/GROND 				&	3.0 h		 	&	$i'$					&	284				&	$1\parcsec26$			&	$19.05\pm0.07$	\\ 
MPG/GROND				&	3.0 h		 	&	$z'$					&	284				&	$1\parcsec39$			&	$18.66\pm0.08$	\\
MPG/GROND 				&	3.0 h		 	&	$J$					&	480				&	$1\parcsec36$			&	$17.84\pm0.09$	\\
MPG/GROND 				&	3.0 h		 	&	$H$					&	480				&	$1\parcsec29$			&	$16.98\pm0.10$	\\
MPG/GROND 				&	3.0 h		 	&	$K_s$				&	480				&	$1\parcsec21$			&	$16.07\pm0.11$	\\ \hline
VLT/HAWK-I				&	355.2 d		&	$J$					&	$240\times10$		&	$0\parcsec6$			&	$22.4\pm0.1$		\\
VLT/HAWK-I				&	355.1 d		&	$K$					&	$240\times10$		&	$0\parcsec5$			&	$20.8\pm0.2$		\\
NOT/ALFOSC				&	483.9 d		&	$B$					&	$5\times480$		&	$1\parcsec3$			&	$24.2\pm0.2$		\\
NOT/ALFOSC				&	475.0 d		&	$R$					&	$9\times265$		&	$1\parcsec1$			&	$23.8\pm0.3$		\\
NOT/ALFOSC				&	440.0 d		&	$i$					&	$9\times330$		&	$0\parcsec9$			&	$23.8\pm0.3$		\\
GTC/OSIRIS				&	491.3 d		&	$g'$					&	$3\times250$		&	$1\parcsec6$			&	$24.9\pm0.1$		\\
GTC/OSIRIS				&	491.3 d		&	$z'$					&	$10\times75$		&	$1\parcsec4$			&	$23.2\pm0.3$		\\
\hline
\end{tabular}
\flushleft{$^a$ Time since the GRB trigger (observer's time frame). \\ For the afterglow measurements time is given in hours, while for the host galaxy, it is shown in days.}
\label{tab:phot}
\end{table*}


\subsection{NOT, GTC and VLT/HAWK-I imaging}
To derive physical parameters of the host of GRB\,121024A via stellar population synthesis modelling, we obtained late-time photometry from VLT/HAWK-I, NOT and GTC. Exposure times and seeing can be found in Table~\ref{tab:phot}.

$J$ and $K$ band images were observed with HAWK-I on the Yepun (VLT-UT4) telescope at the
ESO Paranal Observatory in Chile. HAWK-I is a near-infrared imager with a pixel scale of 0\parcsec106/pix and a total field of view of $7\parcmin5 \times 7\parcmin5$.
$B$, $R$ and $i$ images were obtained with the ALFOSC optical camera on the NOT. The photometric
calibration was carried out by observing the standard star GD71 at a similar airmass to the GRB field.
$g^\prime$ and  $z^\prime$-band host galaxy images were taken with the
10.4m GTC. The images were acquired with the OSIRIS instrument which provides an
unvignetted field of view of $7\parcmin8 \times 7\parcmin8$ and a pixel
scale of 0\parcsec25/pix \citep{cepa00}. Images were taken
following a dithering pattern. The $z^\prime$-band images
were defringed by subtracting an interference pattern which was
constructed based on the dithered individual frames. The photometric
calibration was carried out by observing the standard star SA95-193
\citep{smith02}. NOT and GTC are located at the observatory of
Roque de los Muchachos, La Palma, Spain.

All images were dark-subtracted and flat-fielded using IRAF standard routines.

\subsection{GROND and Skynet Photometry}
GROND data was reduced using standard IRAF tasks \citep{tody97,thomas08}. The afterglow image was fitted using a general point spread function (PSF) model obtained from bright stars in the field. The optical images in $g', r', i', z'$ were calibrated against standard stars in the SDSS catalogue, with an accuracy of $\pm0.03$\,mag. The NIR magnitudes were calibrated using stars of the 2MASS catalogue, with an accuracy of $\pm0.05$\,mag. Skynet obtained images of the field of GRB\,121024A on 2012 October 24-25 with four $16''$ telescopes of the PROMPT array at CTIO, Chile, and the $16''$ DAO in Italy. Exposures ranging from 5 to 160\,s were obtained in the $BVRI$ (PROMPT) and $g', r', i'$ (DAO) bands, starting at 02:57:07\,UT ($t = 55$\,s since the GRB trigger) and continuing until $t = 7.3$\,h on the first night, and continuing from $t = 20.7 - 25.5$\,h on the second night. Bias subtraction and flat-fielding were performed via Skynet's automated pipeline. Post-processing occurred in Skynet's guided analysis pipeline, using both custom and IRAF-derived algorithms. Differential aperture photometry was performed on single and stacked images, with effective exposure times of 5\,s to 20\,min on the first night, and up to $\sim$4\,h on the second night. Photometry was calibrated to the catalogued $B, V, g', r', i'$ magnitudes of five APASS DR7 stars in the field, with $g', r', i'$ magnitudes transformed to RI using transformations obtained from prior observations of Landoldt stars (Henden, A. et al., in preparation). The Skynet magnitudes can be seen in Appendix~\ref{appendix}.

\section{Analysis and Results}\label{results}

\subsection{Absorption Lines}\label{abs}

\begin{figure}
\centering
\resizebox{\hsize}{!}{\includegraphics[bb=29 39 548 358,clip]{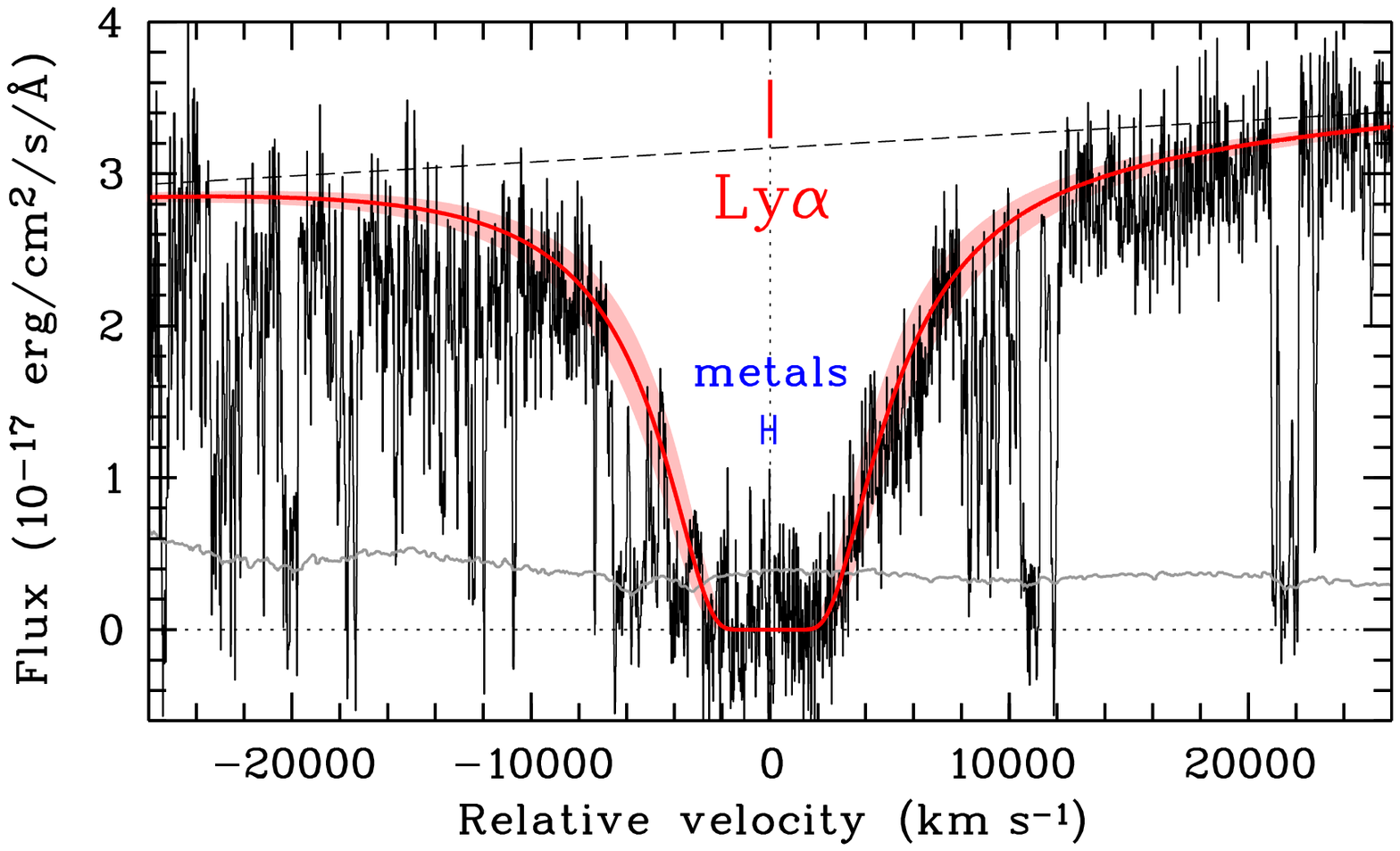}}
\caption{The UVB spectrum centred on the damped Ly$\alpha$ absorption line at the GRB host galaxy redshift. For clarity purposes, the spectrum has been smoothed with a median filter with a sliding window width of 3 pixels. A neutral hydrogen column density fit (log\,$N(\text{H\,I})\,=\,21.88\pm0.10$) to the damped Ly$\alpha$ line is
         shown with a solid line (red), while the 1$\sigma$
         errors are shown with the shaded area (also red). In blue is shown the velocity range of the metal absorption lines. The dashed line shows the continuum placement, while the grey line near the bottom shows the error spectrum.}
\label{fig:dla}
\end{figure}

\begin{table*}
\caption{Ionic column densities of the individual components of the line profile. The transitions used to derive column densities are reported in the second column. Transitions marked in bold are those unblended and unsaturated lines that we use to determine the line-profile decomposition. The results for the ground state and excited levels are listed in the top and bottom part of the table, respectively. Velocities given are with respect to the [\mbox{O\,{\sc iii}}] $\lambda$5007 line ($z=2.3015$). (b) / (s) indicate that the line is blended / saturated. The error on the redshifts of each component is $0.0001$}
\renewcommand*{\arraystretch}{1.2}
\begin{tabular}{@{} l c c c c c c @{}}
\hline\hline
Component	&			Transition								&		a 			&		b			&		c			&		d			&		e 			\\ \hline
$z$			&				---								&		2.2981		&		2.2989		&		2.3017		&		2.3023		&		2.3026		\\
$b$ (km\,s$^{-1}$)&				---								&		26			&		21			&		20			&		22			&		35			\\
$v$ (km\,s$^{-1}$)&				---								&		$-264$		&		$-191$ 		&		$64$    		&		$118$		&		$145$		\\ \hline
log($N$)		&												&					&					&					&					&					\\ \hline
Mg\,{\sc i}		&	$\lambda$1827, $\lambda$2026(b)					&	$13.97\pm0.05$	&	$13.57\pm0.05$	&	$<13.4$			&	$13.57\pm0.07$	&	$<13.4$			\\
Al\,{\sc iii}		&	$\lambda$1854(s), $\lambda$1862(s)				&		---			&		---			&		---			&		---			&		---			\\
Si\,{\sc ii}		&	$\lambda$1808(s)								&		---			&		---			&		---			&		---			&		---			\\
S\,{\sc ii}		&	$\lambda$1253(s)								&		---			&		---			&		---			&		---			&		---			\\
Ca\,{\sc ii}		&	$\lambda$3934, $\lambda$3969					&  $13.25\pm{0.16}^{a}$  &  $12.50\pm{0.16}^{b}$   &	$12.20\pm{0.16}$	&	$11.90\pm{0.16}$	&	$11.20\pm{0.16}$	\\
Cr\,{\sc ii}		&	\textbf{$\boldsymbol{\lambda}$2056}, $\lambda$2062(b), \textbf{$\boldsymbol{\lambda}$2066}	&$13.47\pm0.05$	&	$13.48\pm0.05$	&	$13.39\pm0.05$	&	$13.67\pm0.05$	&	$13.34\pm0.09$	\\
Mn\,{\sc ii}		&	\textbf{$\boldsymbol{\lambda}$2576}, \textbf{$\boldsymbol{\lambda}$2594}, \textbf{$\boldsymbol{\lambda}$2606}	&	$13.15\pm0.05$&	$13.07\pm0.05$&	$12.71\pm0.05$&	$13.21\pm0.05$&	$12.93\pm0.05$	\\
Fe\,{\sc ii} 		&	$\lambda$1611, \textbf{$\boldsymbol{\lambda}$2260}, \textbf{$\boldsymbol{\lambda}$2249}	&$15.15\pm0.05$	&	$15.09\pm0.05$	&	$14.81\pm0.05$	&	$15.27\pm0.05$	&	$15.12\pm0.05$	\\
Ni\,{\sc ii}		&	$\lambda$1345, $\lambda$1454, $\lambda$1467.3, $\lambda$1467.8, $\lambda$1709		&	$13.91\pm0.10$	&	$13.88\pm0.10$	&	$13.95\pm0.09$	&	$14.17\pm0.10$	&	$13.73\pm0.29$	\\
Zn\,{\sc ii}		&	$\lambda$2026(b), $\lambda$2062(b)				&	$13.14\pm0.05$	&	$13.05\pm0.05$		&	$12.50\pm0.08$		&	$13.40\pm0.05$	&	$12.19\pm0.40$	\\		
\hline\hline
Component	&												&	$\alpha$ 			&	$\beta$			&					&					&		 			\\ \hline
$z$			&				---								&	2.2981			&	2.2989			&		---			&		---			&		---			\\
$b$ (km\,s$^{-1}$)&				---								&	28				&	30				&		---			&		---			&		---			\\
$v$ (km\,s$^{-1}$)&				---								&	$-264$   			&	$-191$			&		---			&		---			&		---			\\ \hline
log($N$)		&												&					&					&					&					&					\\ \hline
Fe\,{\sc ii}*	&	$\lambda$2389, $\lambda$2396(b)					&	$13.25\pm0.05$	&	$13.16\pm0.05$	&		---			&		---			&		---			\\
Fe\,{\sc ii}**	&	$\lambda$2396(b), $\lambda$2405(b), $\lambda$2607	&	$12.92\pm0.05$	&	$12.80\pm0.05$	&		---			&		---			&		---			\\
Fe\,{\sc ii}***	&	$\lambda$2405(b), $\lambda$2407, $\lambda$2411(b)	&	$12.63\pm0.05$	&	$12.58\pm0.07$	&		---			&		---			&		---			\\
Fe\,{\sc ii}****	&	$\lambda$2411(b), $\lambda$2414, $\lambda$2622	&	$12.53\pm0.06$	&	$12.61\pm0.05$	&		---			&		---			&		---			\\
Fe\,{\sc ii}*****	&	$\lambda$1559, $\lambda$2360					&	$13.95\pm0.08$	&	$13.68\pm0.13$	&		---			&		---			&		---			\\
Ni\,{\sc ii}**	&	$\lambda$2166, $\lambda$2217, $\lambda$2223		&	$13.43\pm0.05$	&	$13.47\pm0.05$	&		---			&		---			&		---			\\
Si\,{\sc ii}*		&	$\lambda$1309, $\lambda$1533, $\lambda$1816$^{e}$	&$14.98\pm0.11$$^{c}$	&	$14.39\pm0.05^{d}$ &		---			&		---			&		---			\\
\hline
\end{tabular}
\\
\flushleft{
$^{a}$ redshift: 2.2979, $b$-value: 30\,km\,s$^{-1}$, see main text
\\$^{b}$ redshift: 2.2989, $b$-value: 23\,km\,s$^{-1}$, \ \ --
\\$^{c}$ redshift: 2.2979, $b$-value: 20\,km\,s$^{-1}$, \ \ --
\\$^{d}$ redshift: 2.2987, $b$-value: 30\,km\,s$^{-1}$, \ \ --
\\$^{e}$ The column density of the $\alpha$ component of Si\,{\sc ii}* has been determined solely from the $\lambda$ 1816 line.}
\label{tab:components}
\end{table*}

\begin{table*}
\caption{Total column densities (summed among individual velocity components and including excited levels) and abundances with respect to H and Fe.}
\renewcommand*{\arraystretch}{1.3}
\begin{tabular}{@{} l c c c c c c c c @{}}
\hline\hline
Ion		& log($N$/cm$^{-2}$)$_\text{{tot}}$ 	& log($N$/cm$^{-2}$)$_{\text{a+b}}$& log($N$/cm$^{-2}$)$_{\text{c+d+e}}$ & \lbrack X/H\rbrack$_{\text{tot}}$	& \lbrack X/Fe\rbrack&  \lbrack X/Fe\rbrack$_{\text{a+b}}$	& \lbrack X/Fe\rbrack$_{\text{c+d+e}}$\\ \hline
H\,{\sc i}	&		$21.88\pm0.10$		&		---				&		---				&		---				&		---		&		---				&		---				\\
Mg\,{\sc i}	&		$<14.31$				&	$14.11\pm0.03$		&	$<13.86$				&		---				&		---		&		---				&		---				\\
Al\,{\sc iii}	&		$>14.11$				&		---				&		---				&		---				&		---		&		---				&		---				\\
Si\,{\sc ii}	&		$>16.35$				&	 	---				&		---				&		$>-1.0$			&	$>0.53$		&		---				&		---				\\
S\,{\sc ii}	&		$>15.90$				&		---				&		---				&		$>-1.1$			&	$>0.46$		&		---				&		---				\\
Ca\,{\sc ii}	&		$13.37\pm0.12$		&	$13.32\pm0.13$$^{a}$	&	$12.40\pm0.12$		&		$-2.9\pm0.2$		&	$-1.29\pm0.13$&	$-0.97\pm0.14$$^{a}$	&	$-2.02\pm0.11$		\\
Cr\,{\sc ii}	&		$14.18\pm0.03$		&	$13.78\pm0.04$		&	$13.97\pm0.03$		&		$-1.3\pm0.1$		&	$0.22\pm0.05$	&	$0.18\pm0.05$			&	$0.24\pm0.04$			\\
Mn\,{\sc ii}	&		$13.74\pm0.03$		&	$13.41\pm0.04$		&	$13.47\pm0.03$		&		$-1.6\pm0.1$ 		&     $-0.01\pm0.05$ &	$0.03\pm0.05$			&	$-0.04\pm0.04$		\\
Fe\,{\sc ii}	&		$15.82\pm0.05$		&	$15.45\pm0.05$		&	$15.58\pm0.03$		&		$-1.6\pm0.1$		&		---		&		---				&		---				\\
Ni\,{\sc ii}	&		$14.70\pm0.06$		&	$14.33\pm0.05$		&	$14.47\pm0.06$		&		$-1.4\pm0.1$		&	$0.17\pm0.08$	&	$0.02\pm0.08$			&	$0.16\pm0.06$			\\
Zn\,{\sc ii}	&		$13.74\pm0.03$		&	$13.40\pm0.03$		&	$13.47\pm0.04$		&		$-0.7\pm0.1$		&	$0.85\pm0.06$	&	$0.88\pm0.05$			&	$0.83\pm0.05$			\\
\hline
\end{tabular}
\\
\flushleft{
$^a$\,Different a and b broadening parameter and redshift for Ca\,{\sc ii}, see Sect.~\ref{abs}}
\label{tab:metal}
\end{table*}

\begin{figure*}
\includegraphics[width=1.8\columnwidth]{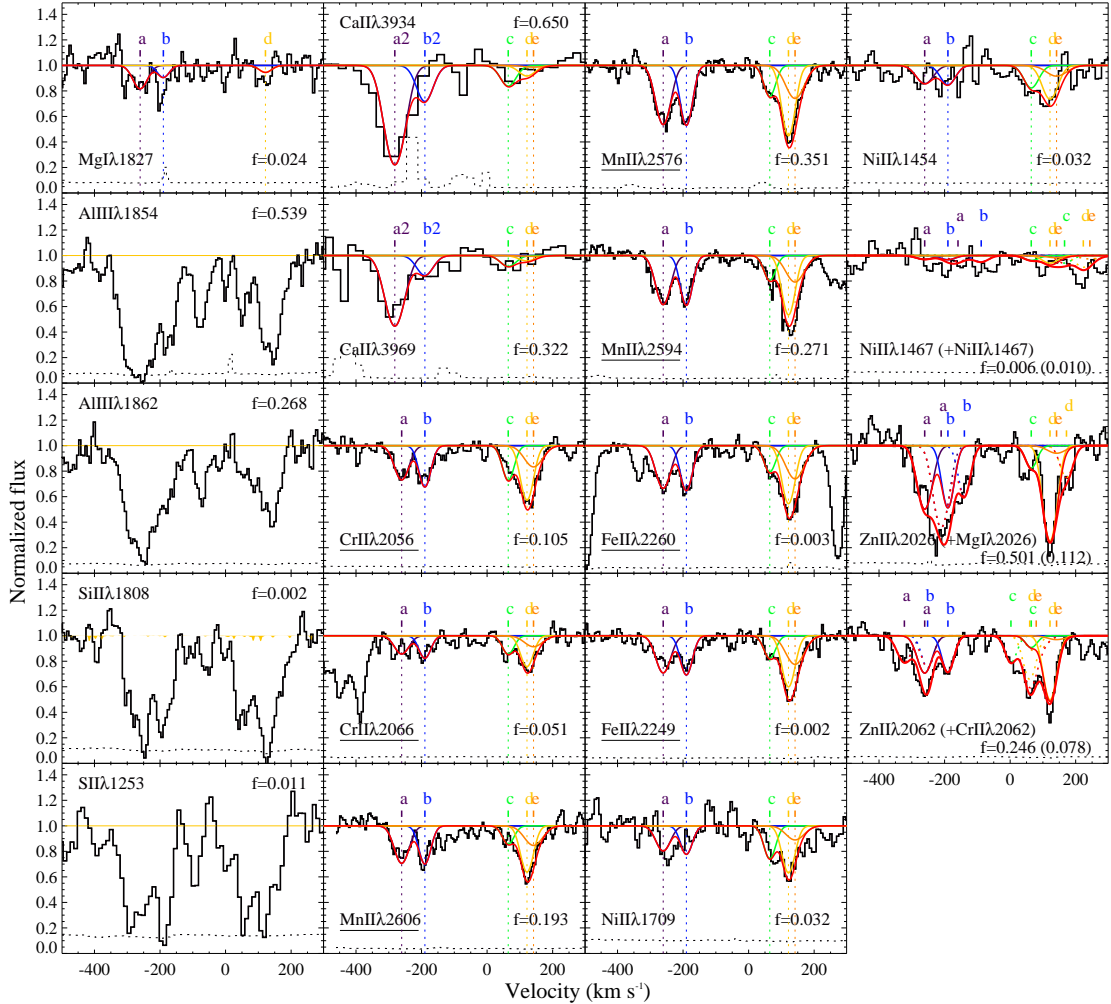}
\caption{Velocity profiles of the metal resonance lines. Black lines show the normalised spectrum, with the associated error indicated by the dashed line at the bottom. The Voigt-profile fit to the lines is marked by the red line, while the single components of the fit are displayed in several colours (vertical dotted lines mark the centre of each component). The decomposition of the line-profile was derived by modelling only the underlined transitions. The oscillator strength 'f', is labelled in each panel. Saturated lines have not been fitted with a Voigt-profile, so for these we show only the spectrum. See online version for colours.}
\label{fig:absorption}
\end{figure*}

\begin{figure*}
\includegraphics[width=1.45\columnwidth]{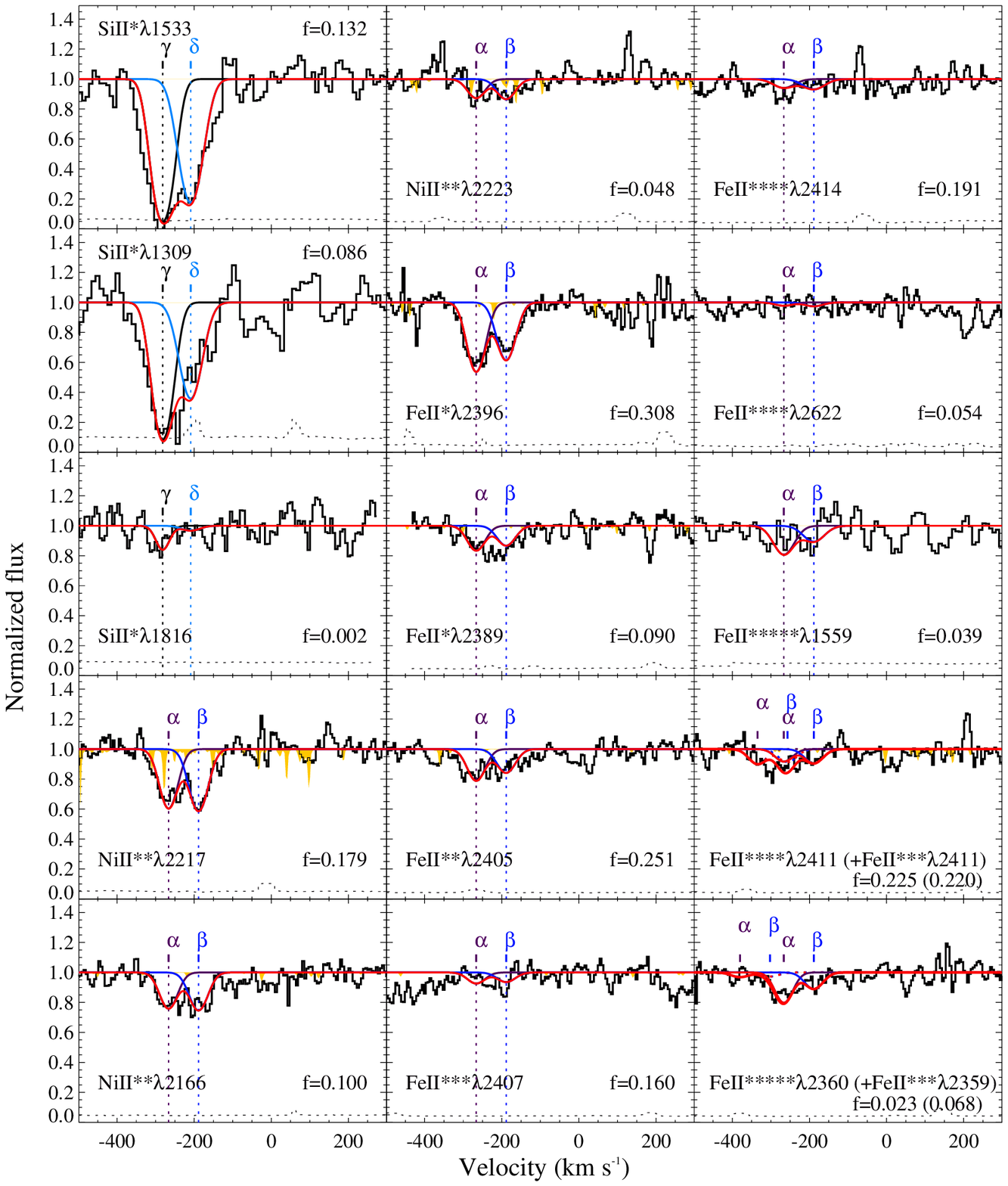}
\caption{The same as Fig.~\ref{fig:absorption}, but for fine-structure lines. Telluric features are highlighted in yellow.}
\label{fig:fine}
\end{figure*}

The most prominent absorption feature is the Ly$\alpha$ line. We plot the spectral region in Fig.~\ref{fig:dla}. Over-plotted is a Voigt-profile fit to the strong Ly$\alpha$ absorption line yielding log\,$N(\text{H\,{\sc i}})=21.88\pm0.10$. The error takes into account the noise in the spectrum, the error on the continuum placement and background subtraction at the core of the saturated lines. Table~\ref{tab:components} shows the metal absorption lines identified in the spectrum.
To determine the ionic column densities of the metals, we model the identified absorption lines with a number of Voigt-profile components, as follows. We use the Voigt-profile fitting software VPFIT\footnote{\url{http://www.ast.cam.ac.uk/~rfc/vpfit.html}} version 9.5 to model the absorption lines. We first normalise the spectrum around each line, fitting featureless regions with zero- or first-order polynomials. To remove the contribution of atmospheric absorption lines from our Voigt-profile fit, we compare the observed spectra to a synthetic telluric spectrum. This telluric spectrum was created following \cite{smette} as described by \cite{annalisa12} and assuming a precipitable water-vapour column of $2.5$\,mm. We systematically reject from the fit the spectral regions affected by telluric features at a level of $>1$ per cent\footnote{This procedure does not aim at reproducing the observed telluric spectrum, but simply reject suspect telluric lines from the Voigt-profile fit.}. None of the absorption lines that we include are severely affected by telluric lines. The resulting column densities are listed in Tables~\ref{tab:components} and~\ref{tab:metal} for lines arising from ground-state and excited levels, respectively. We report formal 1-$\sigma$ errors from the Voigt-profile fitting. We note that these do not include the uncertainty on the continuum normalisation, which can be dominant for weak lines \citep[see e.g.][]{annalisa12}. We hence adopt a minimum error of 0.05\,dex to account for this uncertainty. The error on the redshifts of each component is $0.0001$. The Voigt-profile fits to the metal lines are shown in Figs.~\ref{fig:absorption} and~\ref{fig:fine}.

The fit to the absorption lines from ground-state levels is composed of five components (a-e). We consider the redshift of the [\mbox{O\,{\sc iii}}] $\lambda$5007 emission-line centroid 	$z=2.3015$, as the reference zero-velocity. Components 'a' through 'e' are shifted $-264$, $-191$, $64$, $118$ and $145$\,km\,s$^{-1}$, respectively. Given the resolution of the instrument of 23\,km\,s$^{-1}$ (VIS arm), the individual components are blended, and therefore the profile decomposition is not unequivocal. However, regardless of the properties (and numbers) of the individual components, they are clearly divided into two well separated groups: a+b and c+d+e. When forcing more components to the fit of each group, the resultant total column density are consistent with the previous estimate for each of the two groups. We stress that the resultant $b$-values are not physical, but likely a combination of smaller unresolved components. First we determine redshift $z$ and broadening parameter $b$ (purely turbulent broadening) of the individual components of the line profile, by considering only a master-sample of unblended and unsaturated lines (shown in bold in Table~\ref{tab:components}), with $b$ and $z$ tied among transitions of different ions. Values for $z$ and $b$ were then frozen for the rest of the absorption lines, and the column densities were fitted. We report 3-$\sigma$ lower and upper limits for the saturated and undetected components, respectively. For the saturated lines  Al\,{\sc iii}, Si\,{\sc ii} and S\,{\sc ii} we do not report column densities from the Voigt-profile fit, but instead from the measured equivalent widths (EWs), converted to column densities assuming a linear regime. For these, we only report the total column density for all the components together.

At the H\,{\sc i} column density that we observe, we expect most elements to be predominantly in their singly ionised state \citep{wolfe05}. We hence expect much of the Mg to be in Mg\,{\sc ii} (for this reason we do not report the abundance of Mg\,{\sc i} in Table~\ref{tab:metal}). Ca\,{\sc ii} seems to have a different velocity composition than the rest of the lines. One possibility is that Ca\,{\sc ii} may extend to a slightly different gas phase, as its ionisation potential is the lowest among the observed lines (less than 1\,Ryd = 13.6\,eV). Alternatively, since the Ca\,{\sc ii} lines are located in the NIR arm, a small shift in the wavelength solution with respect to the VIS arm could cause the observed difference. However, a positive comparison between the observed and synthetic telluric lines rules out any shift in the wavelength calibration. We have allowed $z$ and $b$ to have different values for the two Ca\,{\sc ii} lines. This resulted in a slightly different a+b component, but the same c+d+e component as for the rest of the sample.

The fine-structure lines show a different velocity profile composed only of two components, $\alpha$ and $\beta$, see Table~\ref{tab:components}. The redshift of $\alpha$ and $\beta$ are the same as for component a and b found for the resonance lines (but different broadening parameters). Remarkably, no fine-structure lines are detected at the position of components c+d+e. The Si\,{\sc ii}* lines are poorly fitted when tied together with the rest of the fine-structure lines, so we allow their $z$ and $b$ values to vary freely. These components are then referred to as $\gamma$ and $\delta$, which are quite similar to components $\alpha$ and $\beta$, respectively, see Fig.~\ref{fig:absorption}. The column density for component $\gamma$ of the stronger Si\,{\sc ii}* line appears strongly saturated, so only the $\lambda$1816 line has been used to determine the column density in this component. 

The total ionic column densities (summed over individual components and including excited levels when necessary) are given in Table~\ref{tab:metal}. We also report the column densities of the groups of component a+b and c+d+e, which are well resolved from each other, unlike the individual components. Our first metallicity estimate is from Zn, as this element is usually not heavily depleted into dust \citep[see e.g.][]{pettini94}. We derive \lbrack Zn/H\rbrack=$-0.7\pm0.1$ (the other non-refractory elements Si and S are saturated, but the limits we find are consistent). This is in agreement with the value reported in \cite{cucchiara14}.

\begin{figure}
\centering
\includegraphics[width=0.7\columnwidth]{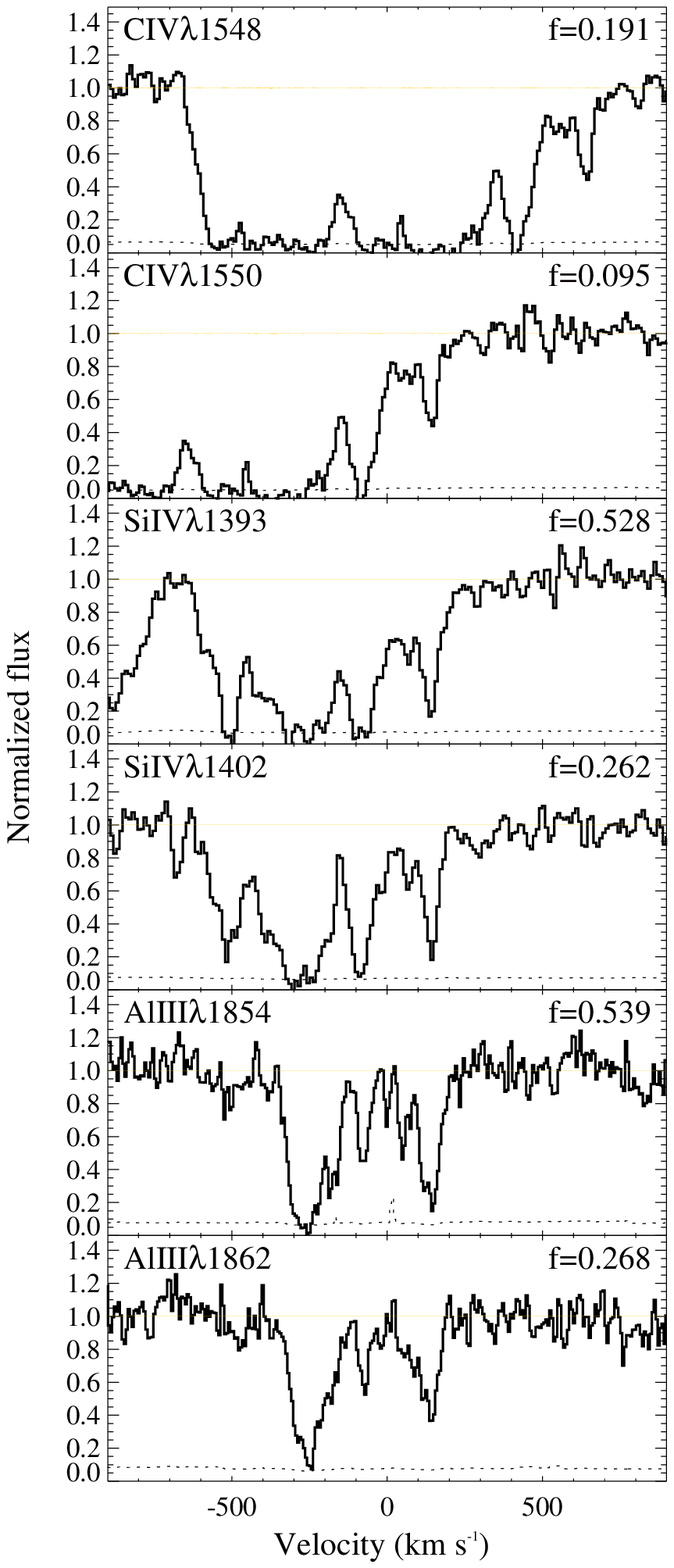}
\caption{High ionisation lines. These lines are highly saturated. See Fig.~\ref{fig:absorption} for details.}
\label{fig:high}
\end{figure}

We note that high ionisation lines from Si\,{\sc iv} as well as C\,{\sc iv} are detected, but are highly saturated, see Fig.~\ref{fig:high}.

\subsection{Dust Depletion}\label{depl}
\label{extinction}
Refractory elements, such as Fe, Ni, and Cr, can be heavily depleted into dust grains \citep[e.g.][De Cia et al. in prep.]{SS96,ledoux02}, and thus can be missing from the gas-phase abundances. A first indicator of the level of depletion in the ISM is the relative abundance [Zn/Fe] (referred to as the depletion factor), because Zn is marginally if not at all depleted into dust grains, and its nucleosynthesis traces Fe. We measure [Zn/Fe] $= 0.85 \pm 0.06$. This value is among the highest for QSO-DLAs, but typical at the observed metallicity of [Zn/H] $=-0.7\pm0.1$ \citep[e.g.][De Cia et al. in prep.]{noterdaeme08}. Following \cite{annalisa13} we calculate a column density of Fe in dust-phase of $\log N(\mbox{Fe})_{\rm dust} = 16.74\pm0.17$ and a dust-corrected metallicity of [Zn/H]$_{\rm corr} = -0.6\pm0.2$, indicating that even Zn is mildly depleted in this absorber, by $\sim0.1$\,dex. This is not surprising given the level of depletion, as also discussed by \cite{jenkins09}.

\begin{figure}
\includegraphics[width=0.95\columnwidth]{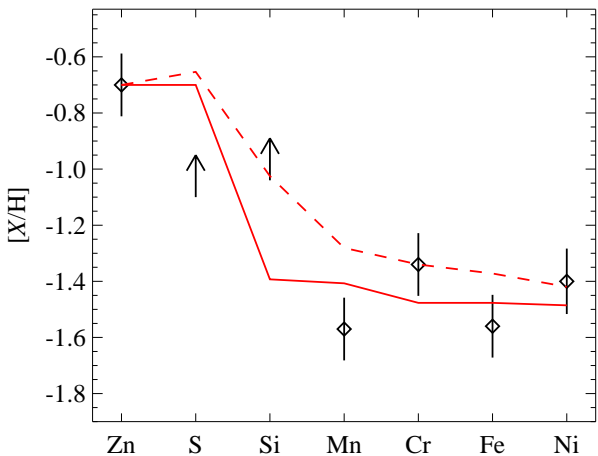}
\caption{The dust-depletion pattern fit for a cold disk (red solid curve) and a warm disk (red dashed curve) to the observed abundances measured from absorption-line spectroscopy (diamonds and arrows, for the constrained and $3\,\sigma$ limits, respectively).}
\label{fig:depletion}
\end{figure}

We also compare the observed abundances of a variety of metals (namely Zn, S, Si, Mn, Cr, Fe, and Ni) to the depletion patterns of a warm halo (H), warm disk+halo (DH), warm disk (WD) and cool disk (CD) types of environments, as defined in \cite{SS96}. These are fixed depletion patterns observed in the Galaxy and calculated assuming that Zn is not depleted into dust grains. We fit the observed abundances to the depletion patterns using the method described in \cite{savaglio01}. We find that none of the environments are completely suitable to describe the observed abundances. The fits to cool- and warm-disk patterns are displayed in Fig.~\ref{fig:depletion} ($\chi^{2}_\nu$=1.18 and 1.58, respectively, with 4 degrees of freedom). For the cool disk the lower limit on the Si column density is not very well reproduced, while the fit for the warm disk overestimates the Mn abundance. The real scenario could be somewhere in between these two environments. Alternatively, the actual depletion pattern is different than what has been observed by \cite{SS96}, or there are some nucleosynthesis effects which we cannot constrain for our case.

Another quantity that is very useful to derive from the observed dust depletion is the dust-to-metals ratio (\dtm{}, normalised by the Galactic value). Constraining the \dtm{} distribution on a variety of environments can indeed shed light on the origin of dust \citep[e.g.][]{mattsson}. Based on the observed [Zn/Fe] and following \cite{annalisa13}, we calculate \dtm{} $=1.01\pm0.03$, i.e. consistent with the Galaxy. From the depletion-pattern fit described above we derive similar, although somewhat smaller, \dtm{} $=0.84\pm0.02$ (CD) and \dtm{} $=0.89\pm0.02$ (WD). These values are in line with the distribution of the \dtm{} with metallicity and metal column densities reported by \cite{annalisa13}, and are also consistent with those of \cite{zafar13}. Following \cite{zafar13}, we calculate \dtm{} $=0.1$ now based on the dust extinction $A_V$ that we model from the SED fit (Sect.~\ref{sed}). Due to the small amount of reddening in the SED, this \dtm{}(A$_V$) value is a factor of 10 lower than expected at the metal column densities observed. This will be discussed further in Sect.~\ref{ext}. 


At the metallicity of GRB\,121024A ($\sim1/3$ solar), it is not possible to draw further conclusions on the dust origin based on the \dtm{}. Both models of pure stellar dust production and those including dust destruction and grain growth in the ISM converge to high (Galactic-like) \dtm{} values at metallicities approaching solar \citep{mattsson}.

\subsection{Distance between GRB and Absorbing Gas}
The most likely origin of the fine-structure transitions observed in the a+b ($\alpha$+$\beta$) component, is photo-excitation by UV photons from the GRB afterglow itself \citep[see e.g][]{prochaska06,vreeswijk07}. Assuming the afterglow to be the only source of excitation, we model the population of the different levels of Fe and Ni, closely following \cite{vreeswijk13}. Using an optical light curve to estimate the luminosity of the afterglow, we can then determine how far the excited gas must be located from the GRB site, for the afterglow to be able to excite these levels. We model the total column density from component a+b ($\alpha$+$\beta$) of all observed levels (ground state and excited states) of  Ni\,{\sc ii} and Fe\,{\sc ii}. We input the optical light curve from Skynet, see Fig.~\ref{fig:skynet} and Tables in Appendix~\ref{appendix}, which is extrapolated to earlier times using the power-law decay observed. We use the broadeningÊparametersÊ$b$ derivedÊfrom the Voigt-profile fits, and the sameÊatomic parameters \citep[see][]{vreeswijk13}.

\begin{figure}
\includegraphics[width=1.20\columnwidth,bb=98 144 592 718]{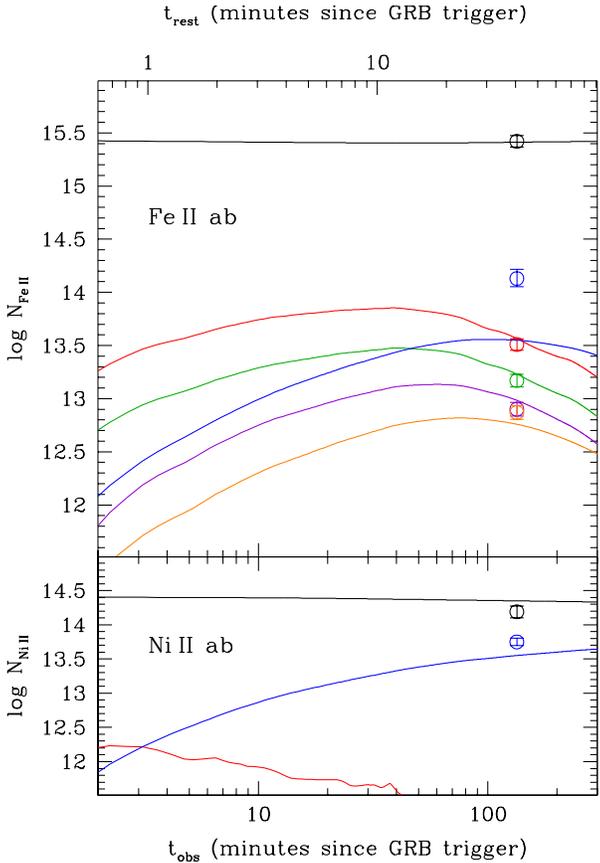}
\caption{Best-fit model for the excited-level populations of the a+b ($\alpha$+$\beta$)
column densities of Fe\,{\sc ii} (top panel) and Ni\,{\sc ii} (bottom). Black lines show the fit to the resonance level. For Fe\,{\sc ii}, from the lower levels and up, excited-level population are shown with red, green, purple, orange and blue. For Ni\,{\sc ii} the red line shows the first excited level, while the blue line shows the second. Open circles show the actual values from Voigt-profile fits. See online version for colours.}
\label{fig:distance}
\end{figure}

The best fit (see Fig.~\ref{fig:distance}) is obtained with a distance of $590\pm100$\,pc between the cloud and burst, and a cloud size of $<333$\,pc (1$\sigma$). The resultant fit is rather poor ($\chi^2$/d.o.f$=40.6/4$). As can be seen in Fig.~\ref{fig:absorption} and~\ref{fig:fine}, the column densities of the ground level of Ni\,{\sc ii} (as probed by Ni\,{\sc ii} $\lambda\lambda\lambda$ 1709, 1454, 1467) and 5th excited level of Fe\,{\sc ii} (as probed by Fe\,{\sc ii} 5s $\lambda\lambda$ 1559, 2360) are not very well constrained due to the observed spectrum having a low S/N ratio near those features. The formal errors from the Voigt profile fit are likely an underestimate of the true error for these column densities. This, in turn, results in the $\chi^2$ of the excitation model fit being overestimated. Furthermore the lack of spectral time series means the resultant parameters are not well constrained. For the c+d+e component we are able to set a lower limit of 1.9\,kpc on the distance to the burst using Fe\,{\sc ii}, and 3.5\,kpc using Si\,{\sc ii} (3$\sigma$). Since Si\,{\sc ii} is saturated, we use the EW to determine the column density, but that only gives the total value of all components together. Hence, for the c+d+e component we fitted using VPFIT and compared the total column density with what we get from the EWs. After establishing that both methods yield the same result, we feel confident in using the column density of log$N$(Si\,{\sc ii})$_{\text{c+d+e}} > 15.99$ together with a detection limit log$N$(Si\,{\sc ii}*)$_{\text{c+d+e}} < 12.80$ on the 1265\,\AA \ line, as this is the strongest of the Si\,{\sc ii}* lines. The lack of vibrationally-excited H$_2$ in the spectra, see below, is in agreement with a distance $\gg100$\,pc, see \cite{draine00}.

\begin{figure}
\includegraphics[bb=48 180 594 612,width=1.10\columnwidth]{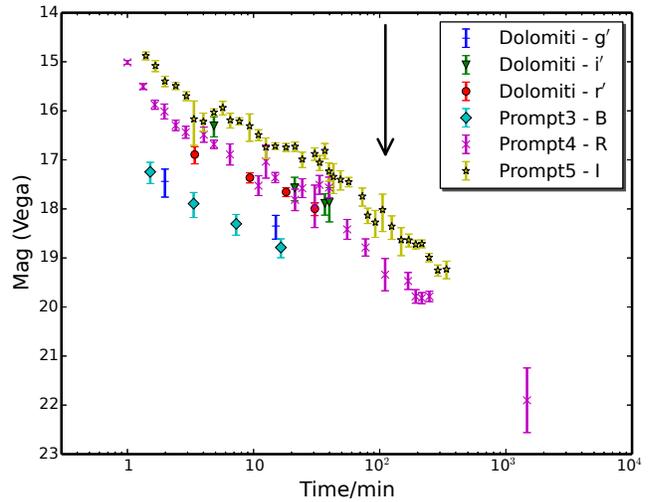}
\caption{GRB-afterglow light curve from the Skynet instruments, used as input for the population modelling. The legend gives the instrument and observational band. The black arrow indicates the starting point of the X-shooter observations. Observations started 55\,s after the GRB trigger. See online version for colours.}
\label{fig:skynet}
\end{figure}

\subsection{Molecular Hydrogen}\label{molecules}

\begin{figure*}
\includegraphics[width=\textwidth]{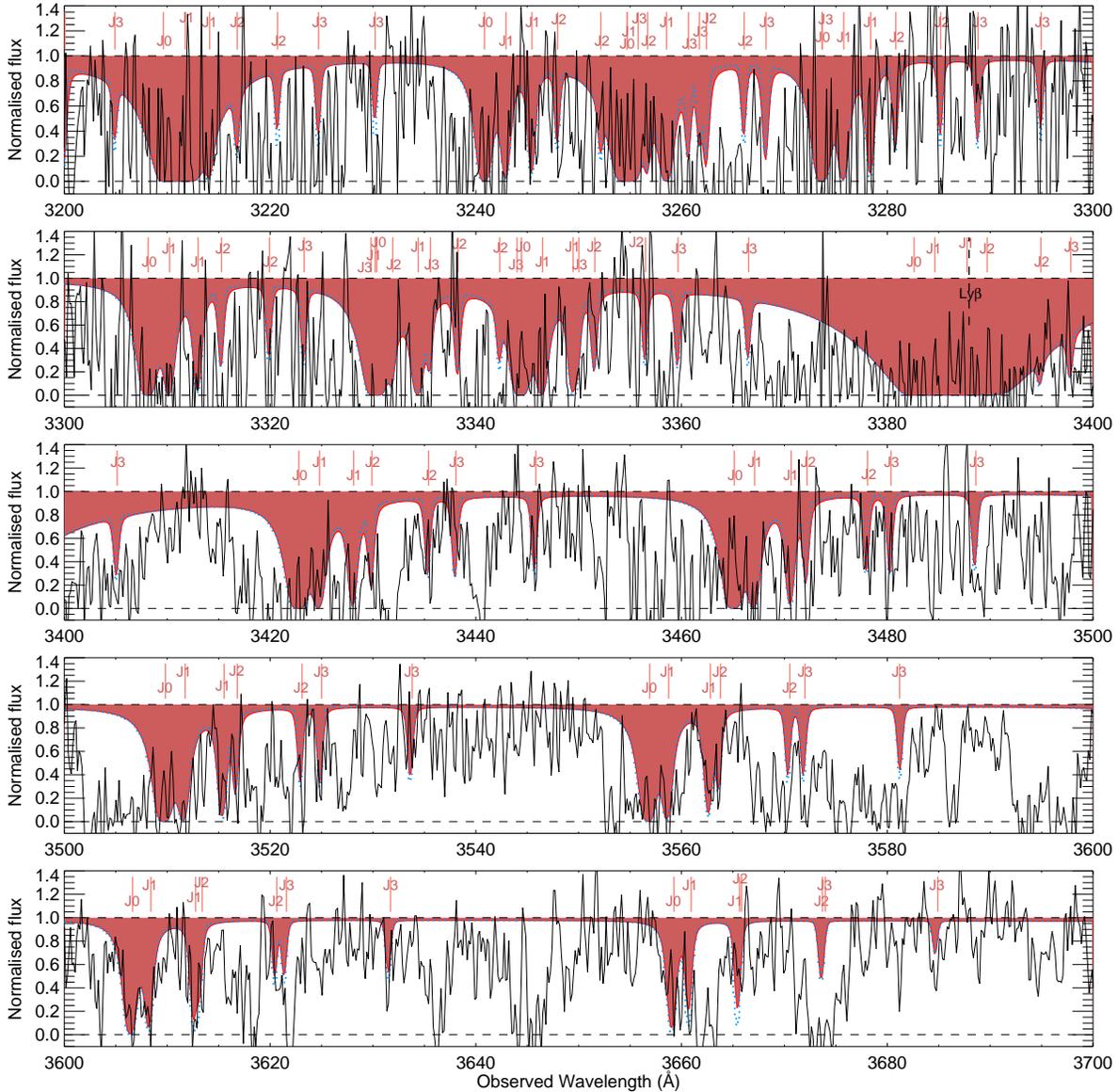}
\caption{X-shooter spectrum showing Lyman- and Werner-band absorption. The shaded area shows the synthetic spectrum from a fit with Doppler parameter $b=1$\,km\,s$^{-1}$, while the blue-dotted line shows the fit for $b=10$\,km\,s$^{-1}$. J0 marks the transitions from the J = 0 rotational level, and likewise for higher J. See online version for colours.}
\label{fig:H2}
\end{figure*}

We detect Lyman- and Werner-band absorption lines of molecular hydrogen at redshift $z\,=\,2.3021$ (corresponding to metal-line component "c+d+e") in rotational levels J=0, 1, 2 and 3, see Fig.~\ref{fig:H2}. The fitting and analysis of the molecular hydrogen transition lines follow \cite{ledoux02,ledoux03} and \cite{thomas1}. We performed a Voigt-profile fit of lines mainly from the Lyman bands L0-0 up to L3-0, as these are found in the less noisy part of the spectrum (a few J\,=\,2 and 3 lines from the Lyman bands L4-0 and L5-0 were also fitted). $J=0$ and 1 lines are strong and fairly well constrained by the presence of residual flux around them, hinting at damping wings in $L\ge 1$. Given the low spectral resolution of the data and the possibility of hidden saturation, we tested a range of Doppler parameters. The estimated H$_2$ column densities, log $N$(H$_2$) are given in Table~\ref{tab:molecules} for Doppler parameter values of $b=1$ and $10$\,km\,s$^{-1}$ resulting in log\,$N(\text{H}_2)$\,=\,19.8--19.9. Using the column density of neutral hydrogen for component 'c+d+e' of log\,$N(\text{H\,I})=21.6$, calculated assuming the same Zn metallicity for the two main velocity components ('a+b' and 'c+d+e'), this results in a molecular fraction in the order of log $f\sim-1.4$, where $f$\,$\equiv$\,2$N$(H$_2$)/($N$(H\,I)+2$N$(H$_2$)). For the component 'a+b' at redshift $z\sim2.2987$, we report log\,$N(\text{H}_2)<18.9$ as a conservative upper limit on detection. A more detailed analysis is not possible because of the high noise-level. The implications of this detection are discussed in Sect.~\ref{GRBmol}. 


We searched for vibrationally-excited H$_2$ by cross-correlating the observed spectrum with a theoretical model from \cite{draine00} and \cite{DH} similar to the procedure outlined in \cite{thomas1}. There is no evidence for H$_{2}^{*}$ in our data, neither through the cross-correlation nor for individual strong transitions, and we set an upper limit of 0.07 times the optical depth of the input model. This approximately corresponds to $\log N$(H$_{2}^{*})<15.7$. A column density of H$_2^*$ as high as seen in e.g. GRBs\,120815A or 080607 \citep{sheffer09} would have been clearly detected in our data.

We furthermore note that CO is not detected. We set a conservative limit of log\,$N(\text{CO})<14.4$, derived by using four out of the six strongest CO AX bandheads with the lowest 6 rotational levels of CO. The wavelength range of the other two bandheads are strongly affected by metal lines, and thus do not provide constraining information.

\begin{table}
\caption{Estimated column densities for H$_2$ for broadening parameter values $b=10$ and $b=1$ (km\,s$^{-1}$).}
\renewcommand*{\arraystretch}{1.3}
\begin{tabular}{@{} l >{\centering\arraybackslash}p{2.5cm} >{\centering\arraybackslash}p{2.5cm} @{}}
\hline\hline
Rotational level	 	& \multicolumn{2}{c}{log($N$(H$_2$)/cm$^{-2}$)}		\\ 
					&	$b=10$\,km\,s$^{-1}$	&	$b=1$\,km\,s$^{-1}$	\\ \hline
$J=0$				& 		19.7				&   	19.7				\\
$J=1$ 				& 		19.2 				&  	19.3				\\
$J=2$				& 		16.1				&   	18.3				\\
$J=3$	 			&		16.0 				&	18.2				\\ \hline
Total					&		19.8				&	19.9				\\
\hline
\end{tabular}
\label{tab:molecules}
\end{table}

\subsection{Emission Lines}
\begin{figure}
\includegraphics[bb=110 162 490 630,width=0.45\columnwidth]{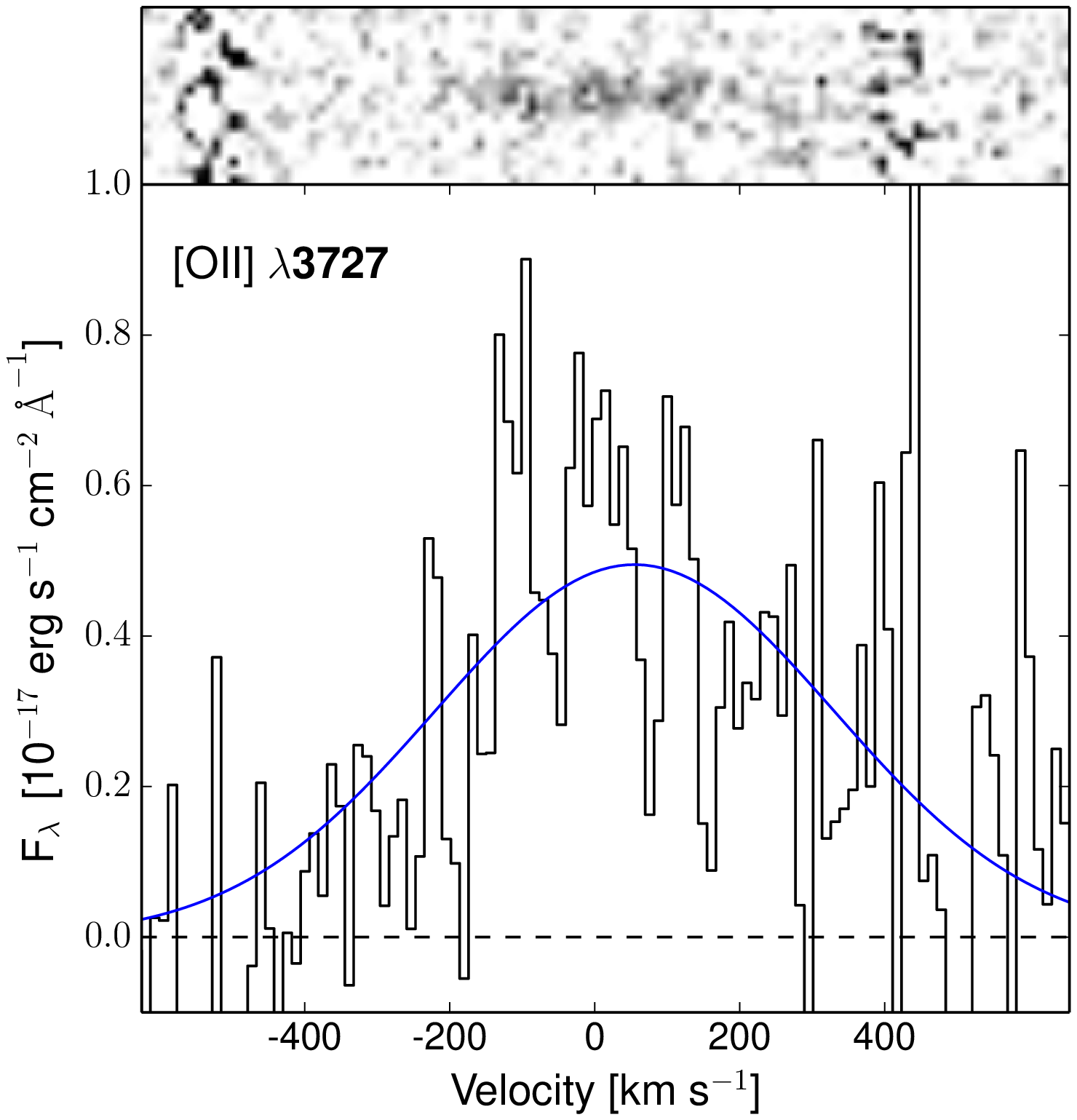}
\includegraphics[bb=110 162 490 630,width=0.45\columnwidth]{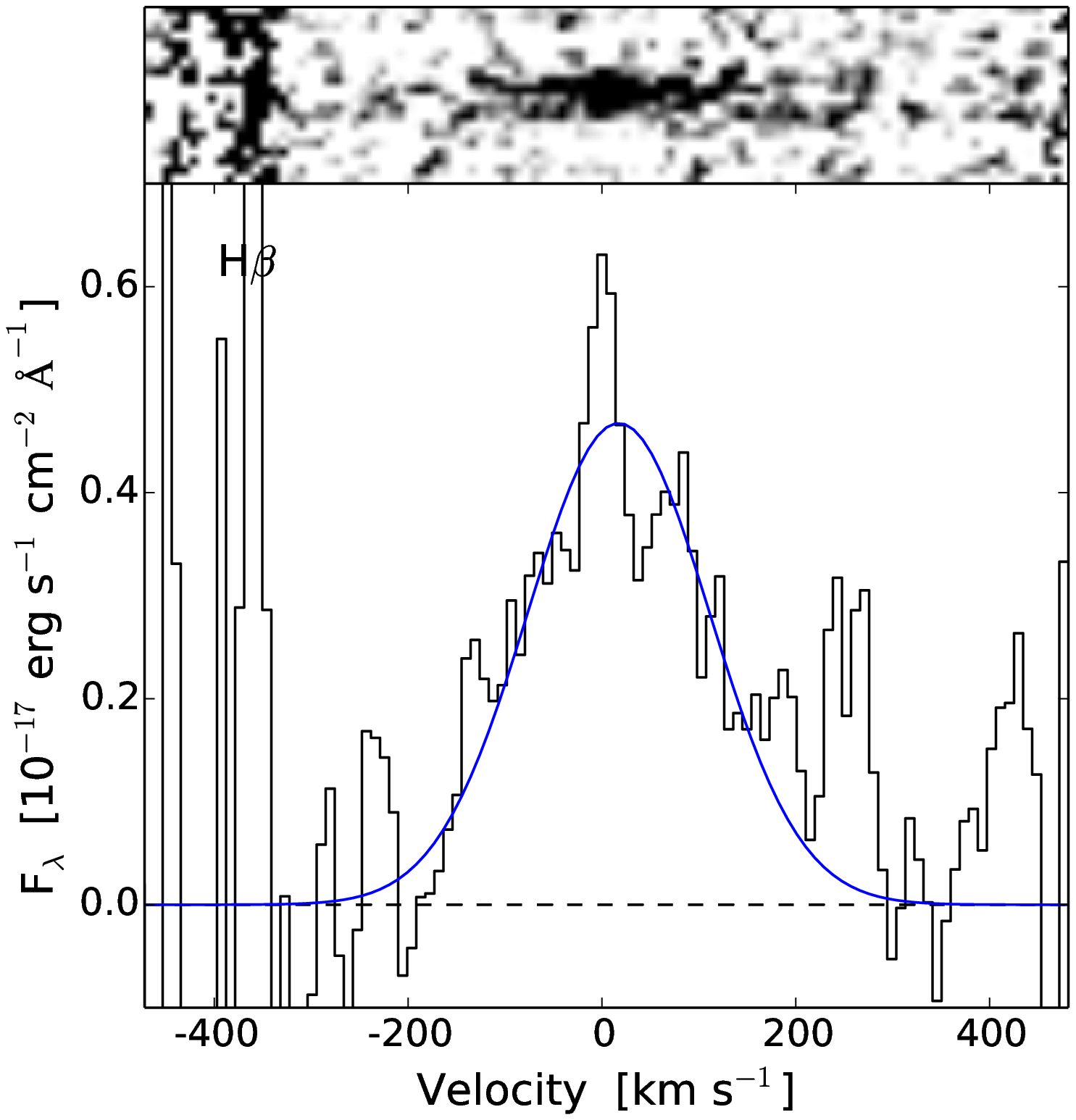}
\includegraphics[bb=110 162 490 630,width=0.45\columnwidth]{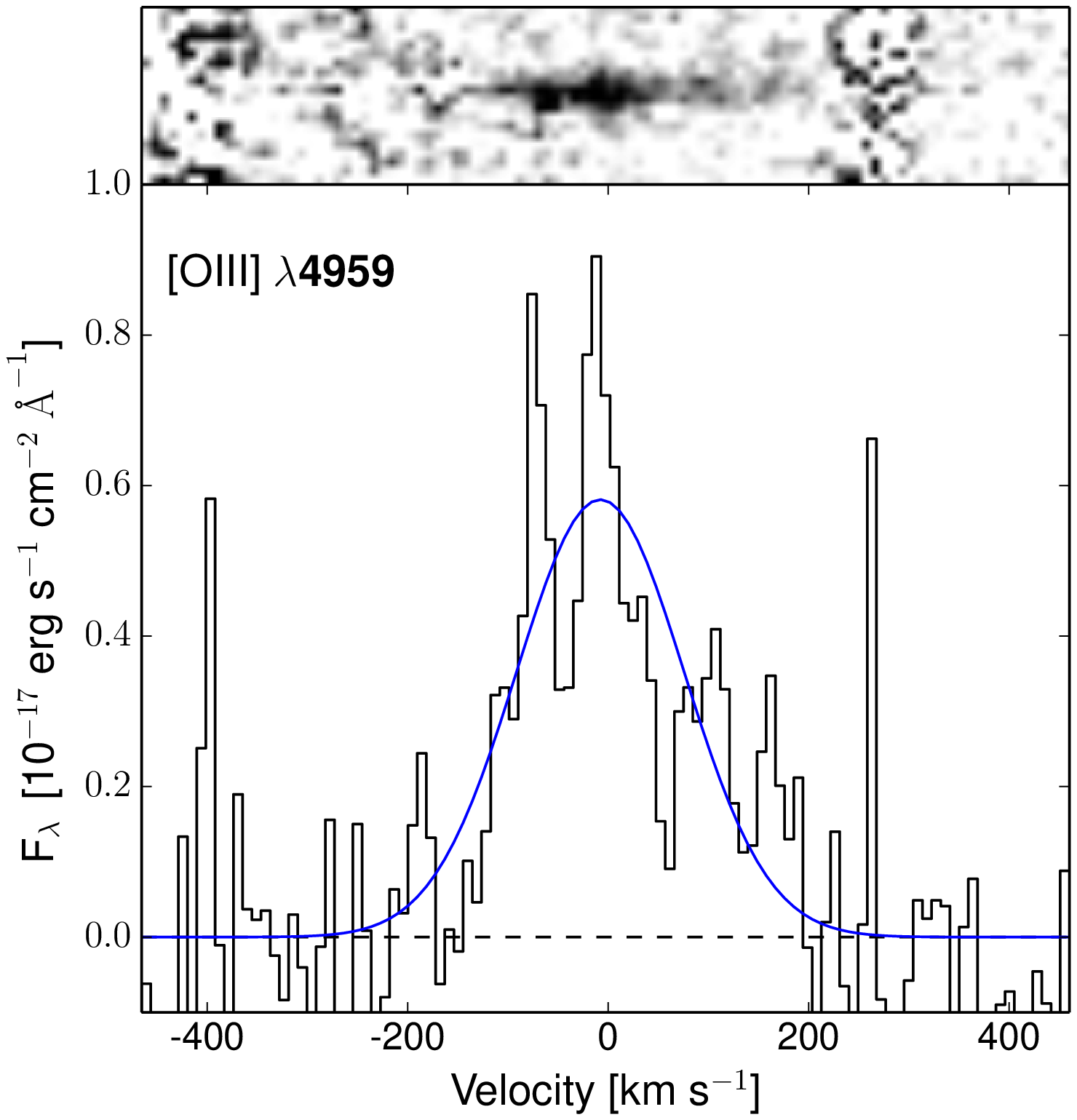}
\includegraphics[bb=110 162 490 630,width=0.45\columnwidth]{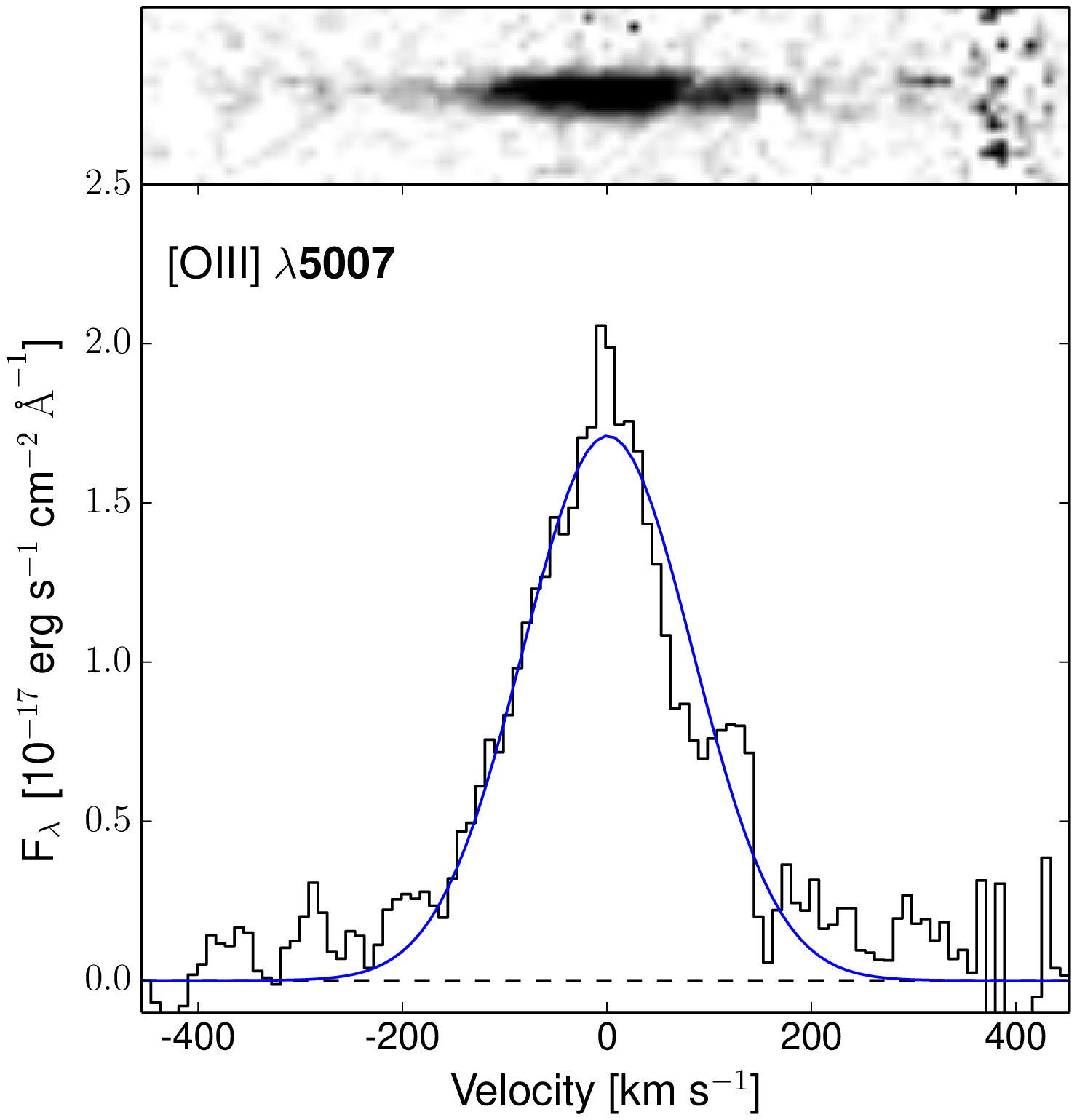}
\includegraphics[bb=110 162 490 630,width=0.45\columnwidth]{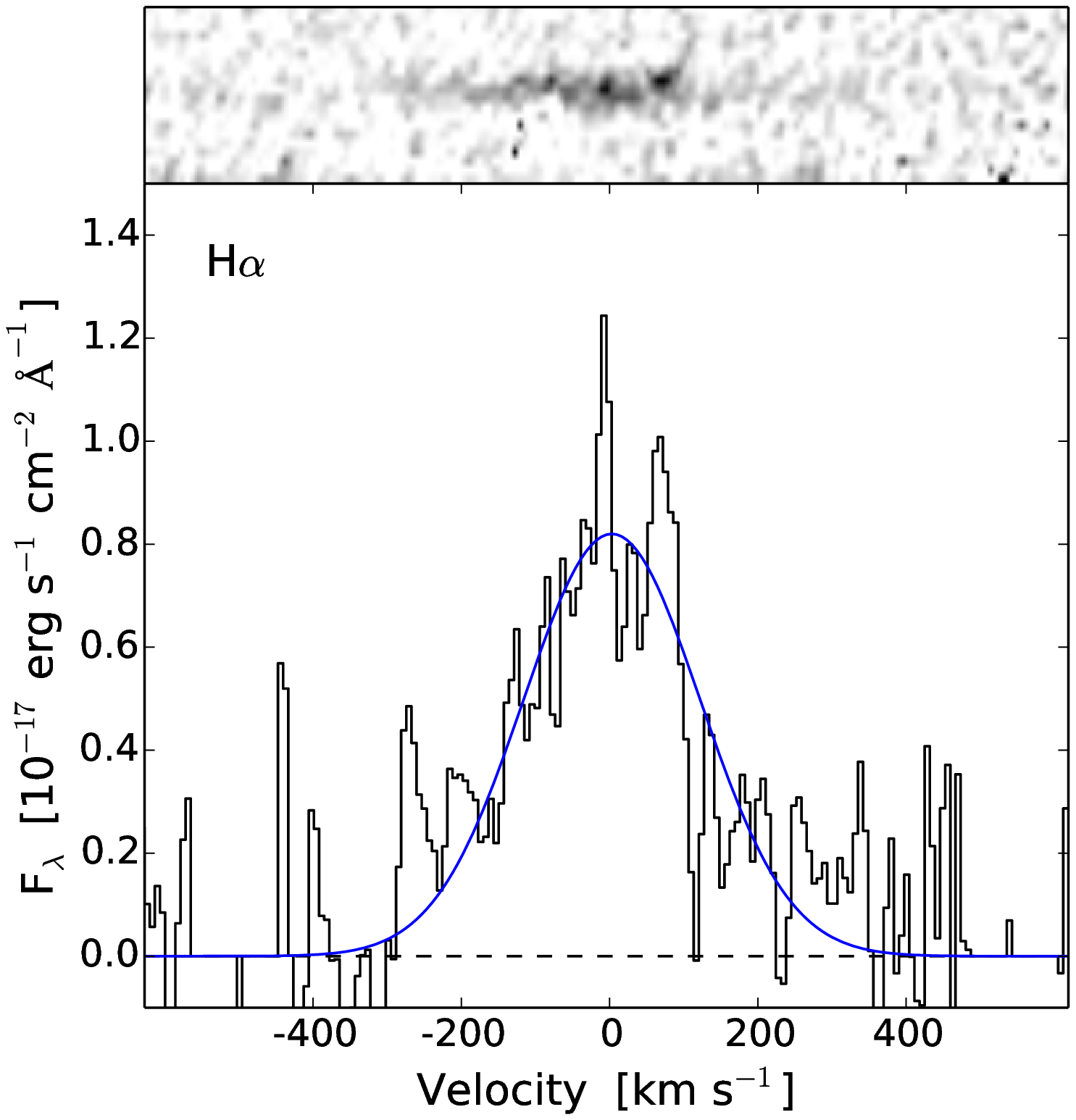}
\includegraphics[bb=-35 4 323 425,width=0.45\columnwidth]{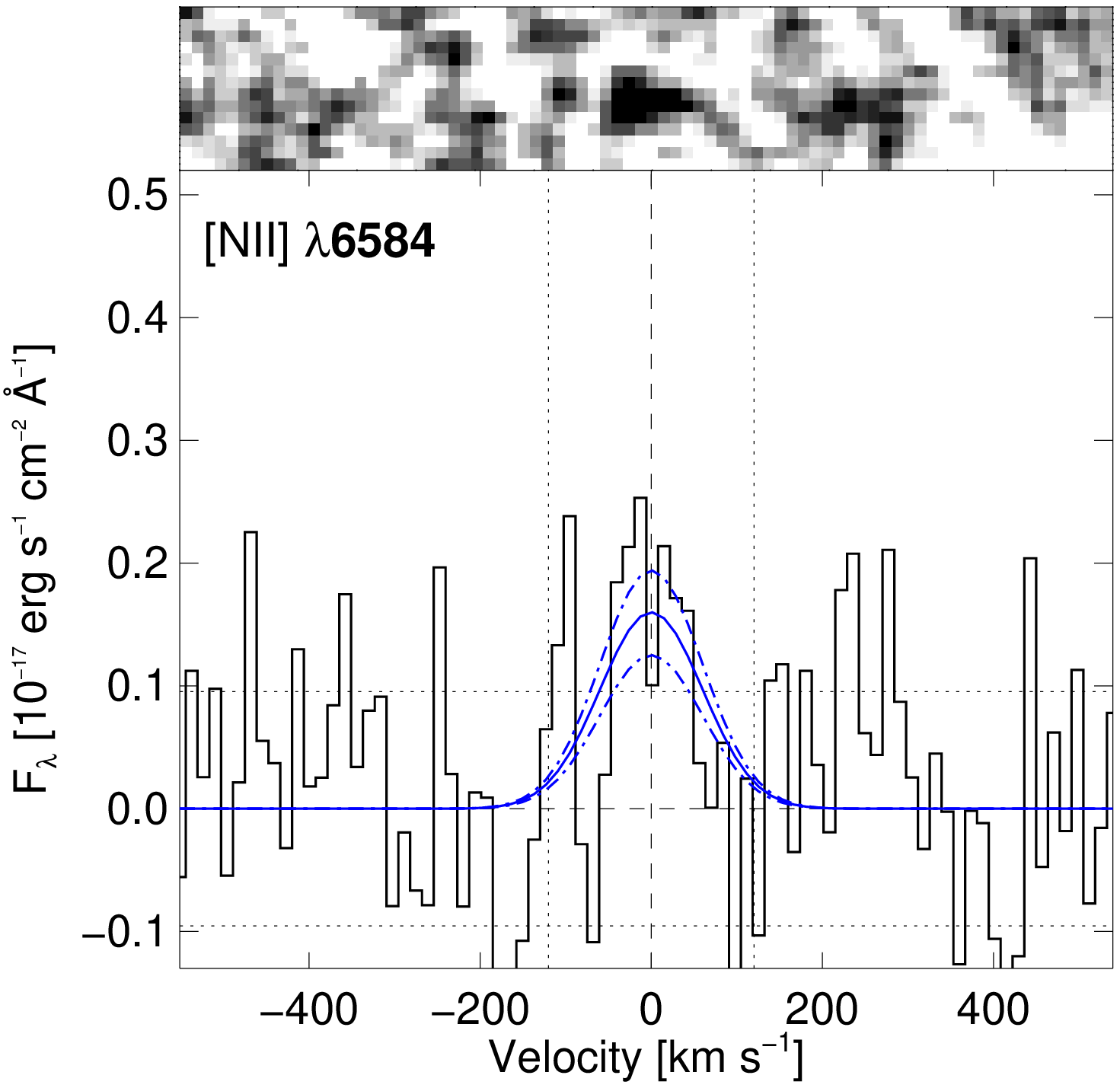}
\caption{Emission lines detected from the GRB\,121024A host. Each panel shows the 2D spectrum after continuum PSF subtraction on top. The bottom part shows the extracted 1D spectrum. The blue line shows the Gaussian fit to the line profile. The abscissa shows the velocity dispersion with respect to the  [\mbox{O\,{\sc iii}}] $\lambda$5007 reference frame. The  [\mbox{N\,{\sc ii}}]  spectrum has been smoothed and binned differently than the other lines, and the fit has been performed with the Gaussian profile centre frozen at 0\,km\,s$^{-1}$ with respect to the reference frame, as indicated with the dashed line in the figure. [\mbox{O\,{\sc ii}}] has been fit as a doublet for the flux estimate.}
\label{fig:emission}
\end{figure}

In the NIR spectrum, we detect H$\alpha$, H$\beta$, the [\mbox{O\,{\sc ii}}] $\lambda$$\lambda$3727, 3729 doublet, [\mbox{N\,{\sc ii}}] $\lambda$6583 (highest-redshift [\mbox{N\,{\sc ii}}] detection published
for a GRB host) and the two [\mbox{O\,{\sc iii}}]  $\lambda$$\lambda$4959, 5007. Table \ref{tab:flux} shows the fluxes (extinction-corrected, see Sect.~\ref{bd}). The reported fluxes are derived from Gaussian fits, with the background tied between the [\mbox{O\,{\sc iii}}] doublet and H$\beta$, and between H$\alpha$ and [\mbox{N\,{\sc ii}}], assuming a slope of the afterglow spectrum of 0.8. [\mbox{O\,{\sc ii}}] is intrinsically a doublet, so we fit a double Gaussian with a fixed wavelength spacing based on the wavelength of the rest-frame lines. Using the GROND photometry, we estimate a slit-loss correction factor of $1.25\pm0.10$.
Fig. \ref{fig:emission} shows the emission-line profiles, the 2D as well as the extracted 1D spectrum. The figure shows a Gaussian fit to the lines, after subtracting the PSF for the continuum \cite[done by fitting the spectral trace and PSF as a function of wavelength locally around each line, see][for details]{moller}. For the weaker [\mbox{N\,{\sc ii}}], a formal $\chi^2$ minimisation is done by varying the scale of a Gaussian with fixed position and width. The noise is estimated above and below the position of the trace (marked by a horizontal dotted line in Fig. \ref{fig:emission}). We assign the zero-velocity reference at the redshift of the [\mbox{O\,{\sc iii}}] $\lambda$5007 line. For the weaker [\mbox{N\,{\sc ii}}] line, we fix the Gaussian-profile fit to be centred at this zero-velocity. 

\begin{table}
\caption{Measured emission-line fluxes}
\renewcommand*{\arraystretch}{1.3}
\begin{tabular}{@{} p{1.1cm} >{\centering\arraybackslash}p{2.2cm} >{\centering\arraybackslash}p{1.1cm} >{\centering\arraybackslash}p{1.1cm} >{\centering\arraybackslash}p{1.3cm} @{}}
\hline\hline
Transition			&	Wavelength$^{a}$	&	Flux$^b$			&	Width$^{c}$	&	Redshift		\\ \hline
[\mbox{O\,{\sc ii}}]  	&	3726.03, 3728.82	&	14.5$\pm$1.2		&	---$^{d}$		&	2.3015$^{e}$	\\
H$\beta$			&	4861.33			&	7.4$\pm$0.4		&	218$\pm$12	&	2.3012		\\
$[\mbox{O\,{\sc iii}}]$&	4958.92			&	9.0$\pm$0.4	  	&	194$\pm$28	&	2.3017		\\
$[\mbox{O\,{\sc iii}}]$&	5006.84			&	27.2$\pm$0.7	 	&	192$\pm$7	&	2.3010		\\
H$\alpha$		&	6562.80			&	21.0$\pm$1.5		&	279$\pm$17	&	2.3010		\\
$[\mbox{N\,{\sc ii}}]$ 	&	6583.41 			&	1.9$\pm$0.7		&	$\sim140$	&	2.3015$^{f}$	\\
\hline
\end{tabular}
\\ $^{a}$ Wavelengths in air in units of \AA.
\\ $^{b}$ Extinction corrected flux in units of 10$^{-17}$\,erg\,s$^{-1}$\,cm$^{-2}$.
\\ $^{c}$ FWHM of line (after removing instrumental broadening) in units of km\,s$^{-1}$. Errors do not include uncertainty in continuum.
\\ $^{d}$ [\mbox{O\,{\sc ii}}] is intrinsically a doublet, which is not fully resolved here, so we do not give the width.
\\ $^{e}$ Calculated using a weighted wavelength average of 3727.7\,\AA.
\\ $^{f}$ The Gaussian fit shown of [\mbox{N\,{\sc ii}}] has a redshift frozen to that of the [\mbox{O\,{\sc iii}}] $\lambda$5007 line.
\label{tab:flux}
\end{table}

\subsection{Star-Formation Rate}\label{sfr}
The SFR can be derived from the emission line fluxes of H$\alpha$ and [\mbox{O\,{\sc ii}}]. Using conversion factors from \cite{kennicutt}, but converted from a Salpeter initial mass function (IMF) to Chabrier \citep{treyer07}, we report extinction corrected (see Sect.~\ref{bd}) values of $\text{SFR}_{H\alpha}=42\pm11$\,M$_\odot$\,yr$^{-1}$ from the H$\alpha$ flux and a $\text{SFR}_{[\mbox{O\,{\sc ii}}]}=53\pm15$\,M$_\odot$\,yr$^{-1}$ derived from [\mbox{O\,{\sc ii}}]. For a comparison with results from the stellar population synthesis modelling see Sect.~\ref{pop}.

\subsection{Metallicity from Emission Lines}
We determine the gas-phase metallicity of the GRB host galaxy using the strong-line diagnostics R$_{23}$ (using the ratio ([\mbox{O\,{\sc ii}}] $\lambda$$\lambda$3727 + [\mbox{O\,{\sc iii}}]  $\lambda$$\lambda$4959, 5007)/H$\beta$), O3N2 (using ([\mbox{O\,{\sc ii}}]/H$\beta$)/(\nitrogen/H$\alpha$)) and N2 \citep[using \nitrogen/H$\alpha$; for a discussion of the different diagnostics see e.\,g.][]{KE}. Note that different metallicity calibrators give different values of metallicity. R$_{23}$ appears to be consistently higher than O3N2 and N2. The R$_{23}$ diagnostic has two branches of solutions, but the degeneracy can be broken using the ratios [\mbox{N\,{\sc ii}}]/H$\alpha$ or [\mbox{N\,{\sc ii}}]/[\mbox{O\,{\sc ii}}]. In our case [\mbox{N\,{\sc ii}}]/H$\alpha$\,=\,0.09$\pm0.02$ and [\mbox{N\,{\sc ii}}]/[\mbox{O\,{\sc ii}}]\,=\,$0.13\pm0.03$, which places the R$_{23}$ solution on the upper branch (though not far from the separation). Because of the large difference in wavelength of the emission lines used for R$_{23}$, this method is sensitive to the uncertainty on the reddening. Both O3N2 and N2 use lines that are close in wavelength, so for these we expect the reddening to have a negligible effect. Instead, they then both depend on the weaker [\mbox{N\,{\sc ii}}] line, which has not got as secure a detection. We derive 12\,+\,log(O/H)\,=\,$8.6\pm0.2$ for R$_{23}$ \citep{mcgaugh91}, 12\,+\,log(O/H)\,=\,$8.2\pm0.2$ for O3N2 and 12\,+\,log(O/H)\,=\,$8.3\pm0.2$ for N2 \citep[both from][]{PP}. The errors include the scatter in the relations \citep[these values are from][and references therein]{KE}, though the scatter in N2 is likely underestimated. See Sect.~\ref{abund} for a comparison with absorption-line metallicity.

\subsection{Balmer Decrement}\label{bd}
The ratio of the Balmer lines H$\alpha$ and H$\beta$ can be used to estimate the dust extinction. We use the intrinsic ratio found I(H$\alpha$)/I(H$\beta)=2.86$ \citep{balmer}, for star-forming regions (and case B recombination, meaning photons above 13.6\,eV are not re-absorbed), where we expect GRBs to occur. The ratio we measure is 2.98 which, assuming the extinction law of \cite{calzetti}\footnote{We use the \cite{calzetti} law, which is an attenuation law for star burst galaxies, where the \cite{pei} laws are relevant for lines of sight towards point-sources inside galaxies where light is lost due to both absorption and scattering out of the line of sight.}, results in $E(B-V)=0.04\pm0.09$\,mag. We note that adopting a different extinction law \citep[from e.g.][]{pei} results in the same reddening correction within errors, because there is little difference within the wavelength range of the Balmer lines.

\subsection{Broad-Band Spectral Energy Distribution}\label{sed}

\begin{figure}
\includegraphics[width=0.7\columnwidth,angle=-90]{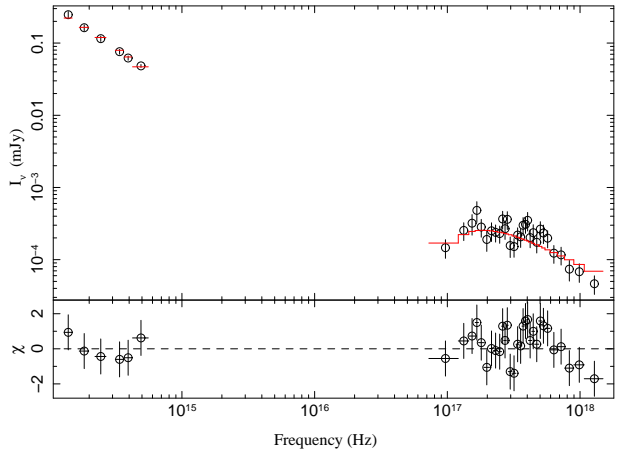}
\caption{NIR-to-X-ray spectral energy distribution and model for the afterglow at 11\,ks after the trigger. The solid red line shows the model. $g'$-band magnitude is not included in the fit, due to possible contribution from the Ly$\alpha$ transition.}
\label{fig:SED}
\end{figure}

We fitted the broad-band afterglow data from XRT and GROND (without the $g'$-band, due to possible DLA contamination), where simultaneous data exist (11\,ks after the trigger). The fit was perform within the ISIS software \citep{isis} following the method of \cite{starling07}. The XRT data were extracted using \emph{Swift} tools. We use single and broken power-law models. For the broken power-law, we tie theÊtwo spectral slopes to a fixed difference of 0.5. Such spectral feature is known as the Òcooling breakÓ of GRB afterglows \citep[e.g.][]{sari98}, and is observed to be the best-fit model for most burst \cite{zafar11}, with the exception of GRB\,080210 \citep{zafar11,annalisa11}. We fit with two absorbers, one Galactic fixed at $N(\text{H})_{\text{X}}^{\text{Gal}}=7.77\times10^{20}$\,cm$^{-2}$ \citep{willingale13}, and one intrinsic to the host galaxy\footnote{We assume solar metallicity, not to provide a physical description of the absorbers, but purely to let $N$(H)$_\text{X}$ conform to the standard solar reference. The reference solar abundances used are from \cite{wilms}.} An SMC dust-extinction model (the average extinction curve observed in the Small Magellanic Cloud) was used for the host, while the reddening from the Milky Way was fixed to $E(B-V)=0.123$ \citep{schlegel}. A single power-law is preferred statistically ($\chi^2$\,/\,d.o.f = 1.07), see Fig.~\ref{fig:SED}, but the two models give similar results.

The best fit parameters for the single power-law SMC absorption model are $N(\text{H})_\text{X}=(1.2^{+0.8}_{-0.6})\times10^{22}$\,cm$^{-2}$ and $E(B-V)=0.03\pm0.02$\,mag at a redshift of $z=2.298$, and a power-law index of $\beta=0.90\pm0.02$ (90 per cent confidence limits), see Table~\ref{tab:sed}. LMC and MW (the average extinction curves observed in the Large Magellanic Cloud and the Milky Way) model fits result in the same values within errors. For a discussion on the extinction see Sect.~\ref{ext}.

\begin{table}
\caption{Best fit parameters from the broad-band spectral energy distribution, for a single power-law SMC absorption model.}
\renewcommand*{\arraystretch}{1.5}
\begin{tabular}{@{} p{4.2cm} >{\centering\arraybackslash}p{3.6cm} @{}}
\hline\hline
$N(\text{H})_\text{X}$					&	$(1.2^{+0.8}_{-0.6})\times10^{22}$\,cm$^{-2}$		\\
$E(B-V)$								&	$0.03\pm0.02$\,mag		\\
Power-law index $\beta$					&	$0.90\pm0.02$		\\
\hline
\end{tabular}
\label{tab:sed}
\end{table}

\subsection{Stellar Population Synthesis Modelling}\label{pop}
Using our photometry of the host, see Table~\ref{tab:phot}, we perform stellar population synthesis modelling of the host galaxy. We use a grid of stellar evolution models with different star formation timescales, age of stellar population and extinction, to compute theoretical magnitudes and compare them to the observed photometry. For the model input, we assume stellar models from \cite{BC03}, based on an IMF from \cite{chabrier03} and a Calzetti dust attenuation law \citep{calzetti}. Table~\ref{tab:pop} lists the galaxy parameters resulting from the best-fit to the HAWK-I, NOT and GTC data. The best fit is obtained with a $\chi^2 = 8$ for the 7 data points used in the modelling. Most of the contribution to the $\chi^2$ comes from the $B$-band observations. This data point lies $\approx$ 3$\sigma$ above the best-fit and the $g$-band measurement, which probes a very similar wavelength range. The reported value of the SFR takes into account the uncertainty in the dust attenuation, and thus has large error bars. We observe a significant Balmer break, which is well fit with star-burst ages between 50 and 500 Myr. The SFR of $\sim40$\,M$_\odot$\,yr$^{-1}$ is consistent with the results from Sect.~\ref{sfr}.


\begin{table}
\caption{Host galaxy parameters from stellar population synthesis modelling}
\renewcommand*{\arraystretch}{1.5}
\begin{tabular}{@{} p{4.2cm} >{\centering\arraybackslash}p{3.6cm} @{}}
\hline\hline
Starburst age (Myr)						&	$\sim250$		\\
Extinction	(mag)						&	$0.15\pm0.15$		\\
M$_\text{B}$							&	$-22.1\pm0.2$		\\
$\rm log(M_*/M_{\odot})$					&	$9.9^{+0.2}_{-0.3}$	\\
SFR (M$_\odot$\,yr$^{-1}$)				&	$40^{+80}_{-25}$	\\
\hline
\end{tabular}
\label{tab:pop}
\end{table}

\subsection{Kinematics}\label{kinematic}
The X-shooter spectrum contains information both on the kinematics of the absorbing gas along the line-of-sight to the location of the burst inside the host galaxy, as well as kinematics of the emitting gas in H\,{\sc ii} regions probed by the emission lines. The emission lines have a full-width-at-half-maximum (FWHM) of around 210\,km\,s$^{-1}$ from a Gaussian fit, see Table~\ref{tab:flux}. We do not observe signs of rotation in the 2 dimensional spectrum. One possibility is that the galaxy could be dominated by velocity dispersion, as observed for galaxies of similar mass and properties, \cite{sins}.
The velocity width that encloses 90\% of the optical depth \citep[as defined by][]{ledoux06} is 460 km s$^{-1}$ based on the Si\,{\sc ii} $\lambda1808$ line. This is consistent with the correlation between absorption line width and metallicity for GRB host galaxies of \cite{arabsalmani}. The velocity for each absorption component, with respect to the emission lines, is given in Table~\ref{tab:components}. The characteristics of different gas components are discussed in Sect.~\ref{gas}.

\subsection{Intervening Systems}\label{intervening}
We identify three intervening systems along the line of sight, at redshifts $z\,=\,2.0798$, $z\,=\,1.959$, and $z\,=\,1.664$. Table~\ref{tab:ew} lists the observed lines along with the measured EWs. Furthermore only at $z\,=\,1.959$ do we observe the Ly$\alpha$ line, but blended with a Si\,{\sc iv} line.
The intervening systems will not be discussed further in this work.


\begin{table}
\caption{Equivalent widths of intervening systems.}
\renewcommand*{\arraystretch}{1.3}
\begin{tabular}{@{} l >{\centering\arraybackslash}p{1.8cm} >{\centering\arraybackslash}p{1.8cm} >{\centering\arraybackslash}p{1.8cm} @{}}
\hline\hline
					 	& \multicolumn{3}{c}{EWs / \AA }									\\ 
Transition					&	$z=2.0798$		&	$z=1.959$		&	$z=1.664$		\\ \hline
C\,{\sc iv}, $\lambda$1548	& 	$0.43\pm0.10$		&   	$3.64\pm0.15$		&	---				\\
C\,{\sc iv}, $\lambda$1550 	& 	$0.38\pm0.10$		&  	$3.11\pm0.14$		&	---				\\
Fe\,{\sc ii}, $\lambda$2382	& 	---				&	$0.72\pm0.10$		&	$0.36\pm0.05$		\\
Mg\,{\sc ii}, $\lambda$2796	&	---	 			&	$3.73\pm0.08$		&	$1.29\pm0.05$		\\
Mg\,{\sc ii}, $\lambda$2803	&	---				&	$2.53\pm0.11$		&	$1.02\pm0.05$		\\
\hline
\end{tabular}
\label{tab:ew}
\end{table}

\section{Discussion and Implications}\label{discussion}
 
\subsection{Abundance Measurements from Absorption and Emission Lines}\label{abund}

The metallicity of GRB hosts is usually determined either directly through absorption line measurements, or via the strong-line diagnostics using nebular-line fluxes. The two methods probe different physical regions; the ISM of the host galaxy along the line of sight as opposed to the ionised star-forming H\,{\sc ii} regions emission - weighted over the whole galaxy. Hence, the two methods are not necessarily expected to yield the same metallicity, see for instance \cite{JW14}. The line of sight towards the GRB is expected to cross star-forming regions in the GRB host. Thus, the absorption and emission lines may probe similar regions. Local measurements from the solar neighbourhood show a concurrence of the two metallicities in the same region, see e.\,g. \cite{esteban04}. Only a few cases where measurements were possible using both methods have been reported for QSO-DLAs \citep[see e.g.][]{bowen05,JW14} but never for GRB-DLAs. The challenge is that the redshift has to be high enough ($z \gtrsim 1.5$) to make Ly$\alpha$ observable from the ground, while at the same time the host has to be massively star-forming to produce sufficiently bright emission lines. Furthermore, the strong-line diagnostics are calibrated at low redshifts, with only few high redshift cases available \citep[see for instance][]{christensen12}. 

The spectrum of GRB\,121024A has an observable Ly$\alpha$ line as well as bright emission lines. We find that the three nebular line diagnostics R$_{23}$, O3N2 and N2 all find a similar oxygen abundance of $12+\text{log(O/H)}=8.4\pm0.4$. Expressing this in solar units we get a metallicity of \lbrack O/H\rbrack\,$\sim-0.3$ (or slightly lower if we disregard the value found from the R$_{23}$ diagnostic, given that we cannot convincingly distinguish between the upper and lower branch). This is indeed consistent with the absorption line measurement from the low-depletion elements (dust-corrected value) [Zn/H]$_{\rm corr}=-0.6\pm0.2$, though the large uncertainty in the strong-line diagnostics hinders a more conclusive comparison.

\cite{krogager13} find a slightly lower metallicity from absorption lines in the spectrum of quasar Q2222-0946, compared to the emission-line metallicity. However, this is easily explained by the very different regions probed by the nebular lines (6\,kpc above the galactic plane for this quasar) and the line of sight, see also \cite{peroux}. QSO lines of sight intersect foreground galaxies at high impact parameters, while the metallicity probed with GRB-DLAs are associated with the GRB host galaxy. Interestingly, \cite{noterdaeme12} find different values for the metallicities, even with a small impact parameter between QSO and absorber, for a QSO-DLAs. A comparison of the two metallicities is also possible for Lyman break galaxies (LBGs), see for instance \cite{pettini02} for a discussion on the metallicity of the galaxy MS 1512--cB58. They find that the two methods agree for a galaxy with an even larger velocity dispersion in the absorbed gas than observed here ($\sim1000$\,km\,s$^{-1}$). The line of sight toward GRB\,121024A crosses different clouds of gas in the host galaxy, as shown by the multiple and diverse components of the absorption-line profiles. The gas associated with component a+b is photo-excited, indicating that it is the closest to the GRB. Given the proximity, the metallicity of this gas could be representative of the GRB birth site. Assuming the GRB exploded in an H\,{\sc ii} region, the emission- and absorption-metallicities are expected to be similar, though if other H\,{\sc ii} region are dominating the brightness, the GRB birth sight might contribute only weakly to the emission line-flux, see Sect~\ref{gas}. Building a sample of dual metallicity measurements will increase our understanding of the metallicity distribution and evolution in galaxies.


\subsection{The Mass-Metallicity Relation at $z\sim2$}
Having determined stellar mass, metallicity and SFR of the GRB host, we can investigate whether the galaxy properties are consistent with the mass-metallicity relation at the observed redshift. Appropriate for a redshift of $z\approx2$, we use equation 5 from \cite{mannucci10}: \\
\\$12 + \text{log(O/H)} = 8.90 + 0.47 \times (\mu_{0.32} - 10)$ \\
\\
where $\mu_{0.32} = \text{log(M}_*[\,\text{M}_\odot]) - 0.32\times\text{log(SFR}_{\text{H}\alpha}[\text{M}_\odot]\,\text{yr}^{-1})$. Using the stellar mass from Sect.~\ref{pop} and the SFR from H$\alpha$ we find an equivalent metallicity of 12\,+\,log(O/H)\,=\,$8.6\pm0.2$ ([O/H]\,=\,$-0.1\pm0.2$). The error does not include a contribution from the scatter in the relation, and is hence likely underestimated. This value is consistent with the metallicity derived from the emission lines, but given the large uncertainty this is perhaps not that illustrative. Instead, we use the mass-metallicity relation determined in \cite{christensen14} for QSO-DLAs for absorption-line metallicities remodelled to GRBs by \cite{arabsalmani}. This results in a metallicity [$M$/H]\,=\,$-0.3\pm0.2$, not including scatter from the relation, and using the mean impact parameter of 2.3\,kpc calculated in \cite{arabsalmani}. This is consistent with the dust-corrected metallicity of [Zn/H]$_{\rm corr}=-0.6\pm0.2$.

\subsection{Grey Dust Extinction?}\label{ext}
We determine the dust extinction/attenuation of the host galaxy of GRB\,121024A both from the Balmer decrement (Sect.~\ref{bd}) and a fit to the X-ray and optical spectral energy distribution (SED, see Sect.~\ref{sed}), as well as from the stellar population synthesis modelling (Sect.~\ref{pop}). The first method determines the attenuation of the host H\,{\sc ii} regions (from the X-shooter spectrum alone), while the SED fitting probes the extinction along the line of sight (using XRT+GROND data). The stellar population synthesis modelling models the host attenuation as a whole (using host photometry). All methods determine the amount of extinction/attenuation by comparing different parts of the spectrum with known/inferred intrinsic ratios, and attribute the observed change in spectral form to dust absorption and scattering. We find values that agree on a colour index $E(B-V)\sim0.04$\,mag. This value is small, but falls within the range observed for GRB-DLA systems. However, low $A_V$'s are typically observed for the lowest metallicities. For our case we would expect a much higher amount of reddening at our determined H\,{\sc i} column density and metallicity. Using the metallicity of [Zn/H]$_{\rm corr}=-0.6\pm0.2$, column density log\,$N(\text{H\,{\sc i}})=21.88\pm0.10$, dust-to-metal ratio \dtm{}$=1.01\pm0.03$ (see Sect. 3.2), and a reference Galactic dust-to-metal ratio $A_{V, \rm{Gal}}/N_{(H, \rm{Gal})}=0.45\times 10^{-21}$ mag cm$^2$ \citep{watson11}, we expect an extinction of $A_V=0.9\pm0.3$\,mag \citep[De Cia et al. in prep. and][]{savaglio03}. This is incompatible with the determined reddening, as it would require $R_V>15$ ($R_V$ for the Galaxy is broadly in the range 2--5). For the Balmer decrement and SED fitting we have examined different extinction curves (MW and LMC besides the SMC) and we have tried fitting the SED with a cooling break, neither option changing the extinction significantly. In an attempt to test how high a fitted reddening we can achieve, we tried fitting the SED with a lower Galactic $N(\text{H\,{\sc i}})$ and reddening. While keeping reasonable values (it is unphysical to expect no Galactic extinction at all), and fitting with the break, the resulting highest colour index is $E(B-V)\sim0.06$\,mag. This is still not compatible with the value derived from the metallicity, so this difference needs to be explained physically.

One possibility to consider is that the host could have a lower dust-to-metals ratio, and hence we overestimate the extinction we expect from the metallicity. However, we see no sign of this from the relative abundances, see Sect.~\ref{depl}. The metallicity is robustly determined from Voigt-profile fits and EW measurement of several lines from different elements (including a lower limit from the none Fe-peak element Si). The lines are clearly observed in the spectra, see Fig.~\ref{fig:absorption}, and the metallicity that we find is consistent with the mass-metallicityÊrelation.

\begin{figure}
\includegraphics[bb=18 180 594 612, width=1.05\columnwidth]{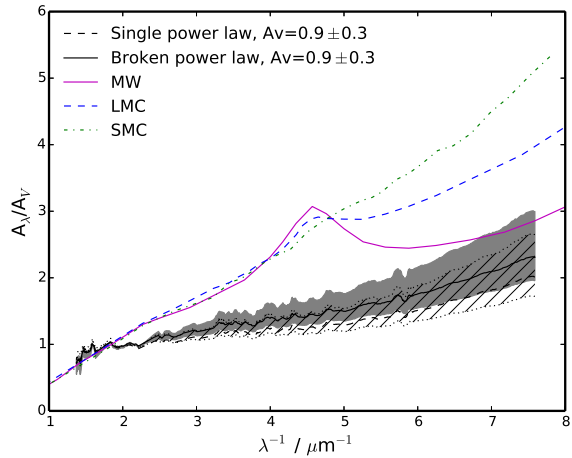}
\caption{Extinction curves for the line of sight to GRB\,121024A. We plot the extinction law assuming $A_V=0.9\pm0.3$\,mag as expected from the measured metallicity, H\,{\sc i} column density and dust-to-metal ratio. The solid black curve shows the extinction curve for a broken power law with $A_V=0.9$, while the grey-shaded area corresponds to the $A_V$ error-space. Likewise, the extinction curve for a single power law is plotted with the dashed black curve, and the hatched area displays the error-space. Over-plotted, in colours, are extinction curves from Pei (1992).}
\label{fig:curves}
\end{figure}

To examine the extinction curve, we perform a fit to the XRT (energy range: 0.3--10\,keV) X-ray data alone and extrapolate the resultant best-fit power-law to optical wavelengths. We try both a single power law, as well as a broken power law with a cooling break in the extrapolation. The latter is generally found to be the best model for GRB extinction in optical fit \citep[e.g.][]{zafar11,greiner11,schady12}. We calculate the range of
allowed $A_\lambda$ by comparing the X-shooter spectrum to the extrapolation, within the 90\,\% confidence limit of the best-fit photon index, and a break in between
the two data sets (X-ray and optical). The resulting extinction does not redden the afterglow strongly, so we cannot constrain the total extinction very well directly from the SED. However, the optical spectroscopy indicates a high metal column density. The strong depletion of metals from the gas phase supports the presence of dust at $A_V=0.9\pm0.3$\,mag (see Sect~\ref{depl}). Fixing $A_V$ at this level, allows us to produce a normalised extinction curve (Fig.~\ref{fig:curves}). This extinction curve is very flat, much flatter than any in the local group \citep{FM}, with an $R_V>9$.

The most likely physical scenario that can explain this shape of the curve is grey dust. If the dust extinction is 'grey', i.e., has a much weaker dependence on wavelength than in the local extinction laws, then a given visual extinction will be much less apparent in the SED ('flat' extinction curve) and thus underestimated in our analysis. Such grey extinction corresponds to larger $R_V$ and is physically interpreted with large grain sizes. A weak wavelength dependance in the extinction for GRBs has been suggested before. \cite{savaglio04} reported a MW-like depletion pattern, but a very low reddening in the SED. \cite{li08} likewise claim a grey extinction law for GRB lines of sight determined by comparing observed spectra to intrinsic ones (by extrapolating from X-rays), arguing for grain growth through coagulation in the dense molecular clouds surrounding GRBs. The larger grains have an extinction that is less dependent on wavelength, because of the contribution of their physical cross-section to the opacity. Preferable destruction of the smaller grains by the GRB would be another possibility, but is unlikely in our case, because the absorbing gas is far from the GRB. We note that the other GRB-DLAs with molecular-hydrogen detection show the expected amount of reddening (using a standard extinction curve), though anomalies do exist in GRB observations. The most notable example to date is reported by \cite{perley08} for GRB\,061126. As for the GRB\,121024A afterglow, they observe a very flat optical--to--X-ray spectral index, arguing for large quantities of grey dust, or a separate origin of the optical and X-ray afterglow. To fit the extinction curve for GRB\,061126, an $R_V\sim10$ is needed. We find an $R_V$ even higher than this, making this case even more extreme than previously observed. We refer to future work on this problem (Friis et al., in prep), as a deeper analysis is beyond the scope of this paper.



\subsection{Molecular Hydrogen in GRB-DLAs}\label{GRBmol}
The lack of detection of molecular hydrogen towards GRBs has puzzled astronomers \citep[see e.\,g][]{tumlinson07}, given that long GRBs are associated with active star formation, and hence are expected to show signatures of molecular clouds. Compared to QSO-DLA line of sights then, we would expect the presence of H$_2$ to be more common for GRB-DLAs, because the QSO-DLA line of sights have a higher probability to intersect the outskirts/halo of the intervening galaxy, where we would anticipate a low molecular content. Recently, a number of H$_2$ detections in GRB afterglows have been reported \citep{prochaska09,thomas1,delia14}, making GRB121024A the fourth definite case. This detection supports the emerging picture that dust has played a major role in biasing past observations against molecular detection \citep[e.g.][]{ledoux09}. Molecules are thought to form on the surface of dust grains, and once formed, shielded from Lyman-Werner photons by the grains. \cite{thomas1} suggest that it is likely this connection that is responsible for the low number of H$_2$ detections towards GRB-DLAs. The high dust column density makes the GRB afterglow UV-faint, preventing high-resolution and high S/N spectroscopy, which is needed to identify the presence of molecular gas. Thus, the lack of H2 detections in most GRB-DLAs can be explained with an observational bias. They illustrate this argument by investigating the metallicity, N(H\,{\sc i}) and dust depletion parameter space, showing that the GRB-DLAs with unsuccessful molecular searches fall outside the region where we would expect detections (with the only exception being GRB\,050829A). This argument is supported by the observed log\,$N(\text{H\,{\sc i}})$, metallicity and depletion factor of GRB\,121024A, which lies inside the parameter space where molecular detections are expect.

The high level of dust depletion observed in this GRB-DLA (see Sect.~\ref{depl}), is consistent with molecular detections in QSO-DLAs \citep{noterdaeme08,thomas1}, where there is a strong preference for H$_2$-bearing DLAs to have significant depletion factors. The dependence on the total neutral hydrogen column density is weak (although intrinsically-weak molecular lines are better constrained in strong DLAs), whereas the parameter that seems to determine whether H$_2$ is detected, is the column density of iron locked into dust. The $\log N({\rm Fe})_{\rm dust}$ that we measure is 2\,dex higher than the column density above which a significant presence of molecules has been observed in QSO-DLAs \citep{noterdaeme08}. Indeed \cite{annalisa13} studied $\log N({\rm Fe})_{\rm dust}$ and concluded that GRB hosts are promising sites for molecular detections.

\cite{delia14} find a molecular fraction for GRB\,120327A of log($f$) between $-7$ and $-4$ with a depletion factor of  \lbrack Zn/Fe\rbrack\,=\,$0.56\,\pm\,0.14$, while for the dustier line of sight towards GRB\,120815A \cite{thomas1} reports a value of log($f$)\,=\,$-1.14\pm0.15$ (\lbrack Zn/Fe\rbrack\,=\,$1.01\,\pm\,0.10$). For GRB\,121024A we find intermediate values, although with the high noise-level the numbers are consistent with those reported for GRB\,120815A. For GRB\,080607 \cite{prochaska09} only report limits on both the molecular fraction and the Zn+Fe column densities. Although the sample is too small to infer anything statistically, it appears that the H$_2$ detection criteria in GRB afterglows follow the trend observed for QSO-DLAs. For a fair estimate of the molecular fraction, the column densities of both H$_2$ and H\,{\sc i} should be constrained for individual velocity components, while this is hardly the case for H\,{\sc i}. Recent work \citep[e.g.][]{balashev15} indicates that the molecular fraction in QSO-DLAs can possibly be much higherÊthan the line-of-sight average values usually measured.

\subsection{Gas kinematics: dissecting the host components}\label{gas}
One of the striking features of the metal absorption-line profiles observed 
towards GRB\,121024A is that they consist of two widely separated groups of 
velocity components (a+b and c+d+e, see Sect.~\ref{abs}). The separation is 
of about 340\,km\,s$^{-1}$, which lies at the high end of the velocity 
distribution of \cite{moller13} and \cite{ledoux06}. The latter is for QSO-DLAs, 
however \cite{arabsalmani} showed that GRB-DLAs follow the velocity-metallicity 
distribution of QSO-DLAs. This suggests that either the two components belong to separate 
galaxies \citep[see for instance][on the GRB\,090323 systems]{savaglio12}, or that this galaxy is fairly 
massive compared to the average GRB host of $\sim$$10^{9}\text{M}_\odot$ 
(\citealt{savaglio09,ceron10}, but see also \citealt{perley13} and 
\citealt{hunt14}). The scenario with separate galaxies is disfavoured, because 
the two absorption components show very similar relative abundances (see 
Table~\ref{tab:components}) and also because the emission lines are centred 
in between the two absorption components \citep[unlike for GRBs\,050820A and 
060418; see][]{hw}. Thus, a likely possibility is that the two absorption 
components are probing different regions within the host. This is in agreement 
with the mass found in Sect.~\ref{pop}, of almost $10^{10}\text{\,M}_\odot$.

Furthermore, the blue (a+b) and the red (c+d+e) absorption components are 
associated with gas at different physical conditions. On one hand, Fe\,{\sc ii}, 
Ni\,{\sc ii} and Si\,{\sc ii} fine-structure lines are detected only in the blue 
component. These lines are photo-excited by the GRB radiation at a distance of 
$\sim$$600$\,pc. On the other hand, H$_2$ molecules are detected in the red 
component only, indicating a gas that is not disturbed by the GRB (at a distance 
of minimum 3.5\,kpc). Through absorption-line spectroscopy at the X-shooter resolution, 
these two different gas components could be located inside the host 
(with respect to the GRB) and characterised.

The observed emission component (arbitrarily set at $v = 0$) traces the brightest 
star-forming regions. Since GRBs tend to reside around the brightest star-forming 
regions in their host \citep{fruchter}, one might expect to observe absorption 
components at velocities close to that of the emission as the line of sight passes 
through this gas. However, the gas around the GRB can be photo-ionised out to 
hundreds of parsecs \citep[e.g.][]{ledoux09,vreeswijk13}; it is thus highly unlikely that 
the optical/UV absorption lines are probing the actual GRB environment. For 
GRB\,121024A, this is further supported by the fact that the a,b component is 
located $\sim$$600$\,pc away from the GRB. Given that giant molecular clouds have a 
maximum radius of $\sim$$200$\,pc \citep{murray11}, the a,b component is undoubtedly 
unrelated to the GRB surroundings.

Although GRBs are most often associated with the brightest star-forming regions, 
this is not always the case. GRB\,980425 \citep[e.g.][]{michal} is an example
where the star-forming region in which the GRB occurred is quite faint compared to 
the larger and brighter star-forming regions in the host. A potentially similar 
scenario could hold for GRB\,121024A as well, in which case the burst should not be 
identified with $v = 0$. In this situation, the possible interpretations of the 
kinematics would be different and lead to other geometric setups compared to those 
conceivable were the GRB localised close to $v = 0$. It should also be noted that we 
have not discussed transverse motion which could complicate the interpretation even 
further. Finally, since the host is most likely an irregular galaxy, indicating a 
3D perturbed environment without a rotating disk, we find it appropriate not to draw 
further conclusions.

While the available data do not allow us to discriminate between possible scenarios, 
this work demonstrates how powerful GRB afterglow observations can be to start 
dissecting individual building-block components of star-forming galaxies at $z\sim2$ 
and above. This is especially true once we have gathered enough data to compile a 
statistical sample; see \citep[e.g.][]{fox08} for previous work on VLT/UVES data and
Fynbo et al. (in prep) for upcoming VLT/X-shooter results on a large afterglow 
sample.

\section*{Acknowledgments}
MF acknowledges support from the University of Iceland Research fund. ADC acknowledges support by the Weizmann Institute of Science Dean of Physics Fellowship and the Koshland Center for Basic Research. The Dark Cosmology Centre is funded by the DNRF. RLCS is supported by a Royal Society Dorothy Hodgkin Fellowship. JPUF acknowledge support from the ERC-StG grant EGGS-278202. CCT is supported by a Ram\'{o}n y Caj\'{a}l fellowship. The research activity of J. Gorosabel, CCT, and AdUP is supported by Spanish research project AYA2012-39362-C02-02. Ad.U.P. acknowledges support by the European Commission under the Marie Curie Career Integration Grant programme (FP7-PEOPLE-2012-CIG 322307). SS acknowledges support from CONICYT-Chile FONDECYT 3140534, Basal-CATA PFB-06/2007, and Project IC120009 "Millennium Institute of Astrophysics (MAS)" of Iniciativa Cient\'{\i}fica Milenio del Ministerio de Econom\'{\i}a, Fomento y Turismo. Part of the funding for GROND (both hardware as well as personnel) was generously granted from the Leibniz-Prize to Prof. G. Hasinger (DFG grant HA 1850/28-1). We thank Alain Smette for providing the telluric spectrum, and the referee for very constructive feedback.

\label{lastpage}

\newpage

\appendix
\section{Skynet magnitude tables}\label{appendix}

Tabels~\ref{tab:skynetR},~\ref{tab:skynetB},~\ref{tab:skynetI},~\ref{tab:skynetgp},~\ref{tab:skynetrp} and~\ref{tab:skynetip} give the magnitudes (not corrected for reddening) used for the optical light-curve input to model the distance between the excited gas (component a+b) and the burst itself.
 
 \newpage
 
\begin{table}
\caption{Skynet - Filter $R$}
\renewcommand*{\arraystretch}{1.3}
\begin{tabular}{@{} l c c c c @{}}
\hline\hline
Filter				&	Time (h)	&	Exposure time		&	S/N		&	Mag. (Vega) 	\\ \hline
$R$				&	0.01656	&	$1\times10$\,s		&	27.7		&	$15.01\pm0.04$	\\ 
$R$ 				&	0.02232 	&	$1\times10$\,s		&	18.9		&	$15.51\pm0.06$	\\ 
$R$ 				&	0.02760	&	$1\times10$\,s		&	12.1		&	$15.88\pm0.09$	\\ 
$R$				&	0.03288	&	$1\times10$\,s		&	7.58		&	$16.0^{+0.2}_{-0.1}$	\\
$R$				&	0.04032	&	$1\times20$\,s		&	10.7		&	$16.3\pm0.1$		\\
$R$				&	0.04824	&	$1\times20$\,s		&	9.35		&	$16.4\pm0.1$		\\
$R$				&	0.06720	&	$1\times40$\,s		&	7.45		&	$16.5^{+0.2}_{-0.1}$	\\
$R$				&	0.08088	&	$1\times40$\,s		&	13.1		&	$16.68^{+0.09}_{-0.08}$	\\
$R$				&	0.10824	&	$1\times40$\,s		&	5.47		&	$16.9\pm0.2$		\\
$R$				&	0.18216	&	$1\times80$\,s		&	5.82		&	$17.5\pm0.2$		\\
$R$				&	0.20880	&	$1\times80$\,s		&	3.58		&	$17.0\pm0.3$		\\
$R$				&	0.24744	&	$1\times160$\,s	&	10.9		&	$17.4\pm0.1$		\\
$R$				&	0.35520	&	$1\times160$\,s	&	5.37		&	$17.8\pm0.2$		\\
$R$				&	0.40536	&	$1\times160$\,s	&	6.08		&	$17.6\pm0.2$		\\
$R$				&	0.50928	&	$1\times160$\,s	&	2.86		&	$17.9^{+0.4}_{-0.3}$	\\	
$R$				&	0.55584	&	$1\times160$\,s	&	6.08		&	$17.5\pm0.2$		\\
$R$				&	0.66000	&	$1\times160$\,s	&	5.53		&	$17.6\pm0.2$		\\
$R$				&	0.91992	&	$3\times160$\,s	&	5.79		&	$18.4\pm0.2$		\\
$R$				&	1.28712	&	$4\times160$\,s	&	6.64		&	$18.79\pm0.2$		\\
$R$				&	1.84008	&	$9\times160$\,s	&	3.68		&	$19.3\pm0.3$		\\
$R$				&	2.79768	&	$7\times160$\,s	&	6.72		&	$19.5^{+0.2}_{-0.1}$		\\
$R$				&	3.21240	&	$7\times160$\,s	&	8.31		&	$19.8\pm0.1$		\\
$R$				&	3.59184	&	$7\times160$\,s	&	10.1		&	$19.8\pm0.1$		\\
$R$				&	4.12824	&	$12\times160$\,s	&	10.7		&	$19.8\pm0.1$		\\
$R$				&	4.9908	&	$18\times160$\,s	&	7.18		&	$20.0\pm0.1$		\\
$R$				&	24.5124	&	$40\times160$\,s	&	1.94		&	$21.9^{+0.7}_{-0.4}$	\\
\hline
\end{tabular}
\label{tab:skynetR}
\end{table}

\begin{table}
\caption{Skynet - Filter $B$}
\renewcommand*{\arraystretch}{1.3}
\begin{tabular}{@{} l c c c c @{}}
\hline\hline
Filter				&	Time (h)	&	Exposure time		&	S/N		&	Mag. (Vega) 	\\ \hline
$B$				&	0.02520	&	$2\times10$\,s		&	4.98		&	$17.2\pm0.2$	\\ 
$B$				&	0.05592	&	$2\times20$\,s		&	4.22		&	$17.9^{+0.3}_{-0.2}$	\\ 
$B$				&	0.12144	&	$3\times40$\,s		&	5.10		&	$18.3\pm0.2$	\\ 
$B$				&	0.27456	&	$2\times160$\,s	&	5.61		&	$18.8\pm0.2$	\\ 
$B$				&	3.87000	&	$37\times160$\,s	&	5.96		&	$21.2\pm0.2$	\\ 
\hline
\end{tabular}
\label{tab:skynetB}
\end{table}

\begin{table}
\caption{Skynet - Filter $I$}
\renewcommand*{\arraystretch}{1.3}
\begin{tabular}{@{} l c c c c @{}}
\hline\hline
Filter				&	Time (h)	&	Exposure time		&	S/N		&	Mag. (Vega) 	\\ \hline
$I$				&	0.02328	&	$1\times5$\,s		&	13.4		&	$14.88\pm0.08$	\\ 
$I$				&	0.02784	&	$1\times5$\,s		&	9.57		&	$15.1\pm0.1$	\\ 
$I$				&	0.03312	&	$1\times10$\,s		&	10.2		&	$15.4\pm0.1$	\\ 
$I$				&	0.04032	&	$1\times20$\,s		&	14.6		&	$15.49^{+0.08}_{-0.07}$	\\ 
$I$				&	0.04896	&	$1\times20$\,s		&	11.8		&	$15.70^{+0.1}_{-0.09}$	\\ 
$I$				&	0.05616	&	$1\times10$\,s		&	2.34		&	$16.2^{+0.5}_{-0.4}$	\\ 
$I$				&	0.06672	&	$1\times40$\,s		&	5.75		&	$16.2\pm0.2$	\\ 
$I$				&	0.08088	&	$1\times40$\,s		&	15.4		&	$16.03\pm0.07$	\\ 
$I$				&	0.09504	&	$1\times40$\,s		&	7.74		&	$15.9\pm0.1$	\\
$I$				&	0.10896	&	$1\times40$\,s		&	7.16		&	$16.2^{+0.2}_{-0.1}$	\\
$I$				&	0.12864	&	$1\times80$\,s		&	15.1		&	$16.21\pm0.07$	\\
$I$				&	0.15504	&	$1\times80$\,s		&	3.84		&	$16.3^{+0.3}_{-0.2}$	\\
$I$				&	0.18216	&	$1\times80$\,s		&	9.82		&	$16.5\pm0.1$	\\
$I$				&	0.20880	&	$1\times80$\,s		&	5.14		&	$16.7\pm0.2$	\\
$I$				&	0.24744	&	$1\times160$\,s	&	16.3		&	$16.72\pm0.07$	\\
$I$				&	0.30144	&	$1\times160$\,s	&	13.3		&	$16.74^{+0.09}_{-0.08}$	\\
$I$				&	0.35520	&	$1\times160$\,s	&	11.3		&	$16.73^{+0.1}_{-0.09}$	\\
$I$				&	0.40536	&	$1\times160$\,s	&	6.89		&	$17.0^{+0.2}_{-0.1}$	\\
$I$				&	0.50928	&	$1\times160$\,s	&	8.51		&	$16.9\pm0.1$	\\
$I$				&	0.55632	&	$1\times160$\,s	&	7.10		&	$17.1^{+0.2}_{-0.1}$	\\
$I$				&	0.61248	&	$1\times160$\,s	&	6.87		&	$16.8^{+0.2}_{-0.1}$	\\
$I$				&	0.66000	&	$1\times160$\,s	&	5.78		&	$17.2\pm0.2$	\\
$I$				&	0.71592	&	$1\times160$\,s	&	3.53		&	$17.3\pm0.3$	\\
$I$				&	0.81528	&	$2\times160$\,s	&	5.50		&	$17.4\pm0.2$	\\
$I$				&	0.94704	&	$2\times160$\,s	&	11.4		&	$17.45^{+0.1}_{-0.09}$	\\
$I$				&	1.21368	&	$1\times160$\,s	&	5.87		&	$17.7\pm0.2$	\\
$I$				&	1.33872	&	$2\times160$\,s	&	6.90		&	$18.1^{+0.2}_{-0.1}$	\\
$I$				&	1.53672	&	$2\times160$\,s	&	3.92		&	$18.3^{+0.3}_{-0.2}$	\\
$I$				&	1.75560	&	$1\times160$\,s	&	2.80		&	$18.0^{+0.4}_{-0.3}$	\\
$I$				&	2.07528	&	$5\times160$\,s	&	4.51		&	$18.4^{+0.3}_{-0.2}$	\\
$I$				&	2.46480	&	$5\times160$\,s	&	3.76		&	$18.6\pm0.3$	\\
$I$				&	2.83080	&	$6\times160$\,s	&	8.36		&	$18.64\pm0.1$	\\
$I$				&	3.21504	&	$7\times160$\,s	&	12.1		&	$18.72\pm0.09$	\\
$I$				&	3.59496	&	$7\times160$\,s	&	13.7		&	$18.71\pm0.08$	\\
$I$				&	4.10328	&	$11\times160$\,s	&	12.8		&	$18.99^{+0.09}_{-0.08}$	\\
$I$				&	4.81200	&	$12\times160$\,s	&	9.54		&	$193\pm0.1$	\\
$I$				&	5.63112	&	$16\times160$\,s	&	6.02		&	$19.2\pm0.2$	\\
$I$				&	24.3192	&	$47\times160$\,s	&	1.89		&	$21^{+0.7}_{-0.4}$	\\
\hline
\end{tabular}
\label{tab:skynetI}
\end{table}

\begin{table}
\caption{Skynet - Filter $g^\prime$}
\renewcommand*{\arraystretch}{1.3}
\begin{tabular}{@{} l c c c c @{}}
\hline\hline
Filter				&	Time (h)	&	Exposure time		&	S/N		&	Mag. (Vega) 	\\ \hline
$g^\prime$			&	0.03312	&	$1\times20$\,s		&	3.76		&	$17.4\pm0.3$	\\ 
$g^\prime$			&	0.25032	&	$1\times80$\,s		&	4.45		&	$18.3^{+0.3}_{-0.2}$	\\ 
$g^\prime$			&	0.69912	&	$1\times80$\,s		&	2.16		&	$19.5^{0.6}_{-0.4}$	\\ 
\hline
\end{tabular}
\label{tab:skynetgp}
\end{table}

\begin{table}
\caption{Skynet - Filter $r^\prime$}
\renewcommand*{\arraystretch}{1.3}
\begin{tabular}{@{} l c c c c @{}}
\hline\hline
Filter				&	Time (h)	&	Exposure time		&	S/N		&	Mag. (Vega) 	\\ \hline
$r^\prime$			&	0.05688	&	$1\times20$\,s		&	6.15		&	$16.9\pm0.2$	\\ 
$r^\prime$			&	0.15576	&	$1\times80$\,s		&	10.6		&	$17.4\pm0.1$	\\ 
$r^\prime$			&	0.30168	&	$1\times160$\,s	&	11.8		&	$17.66^{+0.1}_{-0.09}$	\\ 
$r^\prime$			&	0.50952	&	$1\times160$\,s	&	8.21		&	$18.0\pm0.1$	\\ 
$r^\prime$			&	0.56088	&	$1\times80$\,s		&	5.14		&	$18.0\pm0.2$	\\ 
\hline
\end{tabular}
\label{tab:skynetrp}
\end{table}

\end{document}